\documentclass[10pt,a4paper,aps,twocolumn,showpacs,amssymb,amsmath, color,floatfix]{revtex4-1}
\usepackage{ucs}
\usepackage{amsmath}
\usepackage{amsfonts}
\usepackage{amssymb}
\usepackage{graphicx}
\graphicspath{{EPS_figs/}{PDF_figs/}}
\usepackage{color}

\begin{document}
\title{Theory of elastic interaction between colloidal particles in the nematic cell in the presence of the external electric or magnetic field}
\author{S. B. Chernyshuk $^{1)}$, O.M. Tovkach $^{2)}$ and B. I. Lev $^{2)}$}
\affiliation{ $^{1)}$ Institute of Physics, NAS Ukraine, Prospekt
Nauki 46, Kyiv 03650, Ukraine}
\affiliation{ $^{2)}$ Bogolyubov Institute of
Theoretical Physics,NAS Ukraine,Metrologichna 14-b, Kyiv 03680,
Ukraine.}
\begin{abstract}
The Green function method developed in Ref.[S.~B.~Chernyshuk and B.~I.~Lev, Phys. Rev. E \textbf{81}, 041707 (2010)] is used to describe elastic interactions between axially symmetric colloidal particles in the nematic cell in the presence of the external electric or magnetic field. General formulas for dipole-dipole, dipole-quadrupole and quadrupole-quadrupole interactions in the homeotropic and planar nematic cells with parallel and perpendicular field orientations are obtained. A set of new results has been predicted: 1) \textit{Deconfinement effect} for dipole particles in the homeotropic nematic cell with negative dielectric  anisotropy $\Delta\varepsilon<0$ and perpendicular to the cell electric field, when electric field is approaching it's Frederiks threshold value $E\Rightarrow E_{c}$. This means cancellation of the confinement effect found in Ref. [M.Vilfan et al. Phys.Rev.Lett. {\bf 101}, 237801, (2008)] for dipole particles near the Frederiks transition while it remains for quadrupole particles. 2) New effect of \textit{attraction and stabilization} of the particles along the electric field parallel to the cell planes in the homeotropic nematic cell with $\Delta\varepsilon<0$ . The minimun distance between two particles depends on the strength of the field and can be ordinary for . 3) Attraction and repulsion zones for all elastic interactions are changed dramatically under the action of the external field.

\end{abstract}

\maketitle

\section{Introduction}

Colloidal particles in nematic liquid crystals (NLC) have attracted a great amount of research interest over the last decade. Anisotropic properties of the host fluid - liquid crystal - give rise to a new class of colloidal anisotropic interactions that never occur in isotropic hosts. The world of anisotropic  liquid crystal colloids is much more varied than that of isotropic liquids. 

Experimental study of anisotropic colloidal interactions in the bulk NLC has been made in \cite{po1}-\cite{jap2}. These interactions result in different structures of colloidal particles, such as linear chains along the director field, for particles with dipole symmetry of the director, \cite{po1,po2} and inclined chains with respect to the director for quadrupole particles \cite{po2}-\cite{kot}. Colloidal particles suspended at the nematic-air interface form 2D hexagonal structures \cite{nych,R10}. That structures can be photochemically switched with help of laser light \cite{we_light}.
Quasi two-dimensional nematic colloids in thin nematic cells form a rich variety of 2D crystals by using laser tweezers manipulations. There are 2D hexagonal quadrupole crystals \cite{Mus,s1}, anti-ferroelectric dipole type 2D crystals \cite{Mus,s2} and mixed 2D crystals \cite{ulyana} sandwiched between cell walls. 2D colloidal crystals assembled from chiral colloidal dimers in twisted nematic cell was found in \cite{vortex}. Long ranged elastic interactions between colloids have been experimentally found to be exponentially screened (confinement effect) in the nematic cell across distances compared to the cell thickness $L$ \cite{conf}.

Experimental results are reproduced by using the Landau - de Gennes free-energy numerical minimization approach \cite{Mus},\cite{conf}-\cite{stark2} as well as molecular dynamics \cite{andr1}. 

Theoretical understanding of the matter in the bulk NLC is based on the multipole expansion of the director field and has deep electrostatic analogies \cite{lupe}-\cite{tas1}. In spite of some differences in these approaches, only one of them \cite{lupe} gives exact analytical quantitative result which has been proven experimentally in unlimited nematic LC. Authors of \cite{noel} measured directly the dipole-dipole interaction of iron spherical particles in a magnetic field and found it to be in accordance with \cite{lupe}, within a few percents accuracy for a dipole term. Authors of \cite{jap} and \cite{jap2} have measured experimentally dipole-dipole interaction and found it to be in accordance with  \cite{lupe} within about $10\%$ accuracy. This allows to justify main assumptions of \cite{lupe} for spherical particles in infinite nematic liquid crystal. 

Recently a proposal was made \cite{we,we2} to expand the description of colloidal particles on the confined NLC, which may be a generalization of the method \cite{lupe} for the bulk nematic LC. Using that approach elastic interactions between colloidal particles were found in the nematic cell and near one wall with either planar or homeotropic boundary conditions. The proposed theory \cite{we2} fits very well with experimental data for the confinement effect of elastic interaction between spheres in the homeotropic cell taken from \cite{conf} in the range $1\textendash1000 kT$.

Experimentally influence of the external fields on the nematic colloids were performed in papers \cite{lavr3}-\cite{lopo}. In \cite{lavr3}  bidirectional motion of colloidal particles was observed in nematic cell under the action of ac electric field. Electrically driven multiaxis rotations of colloids was found in \cite{smal1}. Transformation of the 2D colloidal hexagonal lattice at the nematic interface into the chains under the action of the magnetic field was found in \cite{we_magf}. Electrically driven transformations of the Saturn ring director configuration around the spherical particle into point hedgehog was experimentally observed in \cite{lopo} and theoretically predicted in \cite{stark1}.
Qualitative investigation of the magnetic field influence on the elastic quadrupole interaction was made in \cite{ter1}.

Nevertheless there were no effective theoretical approaches for quantitative treatment of the elastic colloidal interactions in the presence of the electric (magnetic) fields. In this paper we propose the Green function method which enables to describe quantitatively elastic colloidal interactions between axially symmetric particles confined in nematic cell under the action of the external electric (magnetic) fields. The current method is the generalization of the method proposed in \cite{we,we2} for description of the colloids in confined NLC. 

The outline of the paper is the following: In Sec. II we present the general Green function method for the description of colloidal particles in confined NLC in the presence of the external electric (magnetic) field. As an illustration of the method we find elastic dipole-dipole, dipole-quadrupole and quadrupole-quadrupole interactions in the unlimited bulk NLC in the presence of the electric field. We draw an analogy with previous results obtained in \cite{we2} for planar nematic cell and propose a correspondence between electric field in unlimited NLC and nematic cell without any field. From this correspondence we propose a criterion for stability of the Saturn ring vs point hedgehog configurations around spherical particle in the nematic cell.  Sec. III presents the energy of elastic interactions in the homeotropic cell placed in the external field. We consider cases of positive $\Delta\varepsilon>0$ and negative $\Delta\varepsilon<0$ NLC dielectric anisotropy and find that they are different in essence. The $\Delta\varepsilon<0$ case lead to the \textit{deconfinement effect} for dipole-dipole interaction when electric field approaches the Frederiks threshold value $E\Rightarrow E_{t}$. Deconfinement effect is the cancellation of the confinement effect which found in \cite{conf} for quadrupole interaction. In this section Sec. III we find as well new effect of \textit{attraction and stabilization} of the particles along the electric field $\textbf{E}$ parallel to the cell planes in the homeotropic nematic cell for $\Delta\varepsilon<0$ case.

In Sec. IV we present the energy of elastic interactions in the planar nematic cell placed in the external field. Here three possible orientations of the field are considered ($\textbf{E}||x,y,z$ axis). Significant changes of attraction and repulsion zones takes place in this case. And Sec. V represents conclusions.

\section{Effective functional for the colloidal nematic cell placed in an external field}

\begin{figure}
\begin{center} 
\includegraphics[width=\columnwidth]{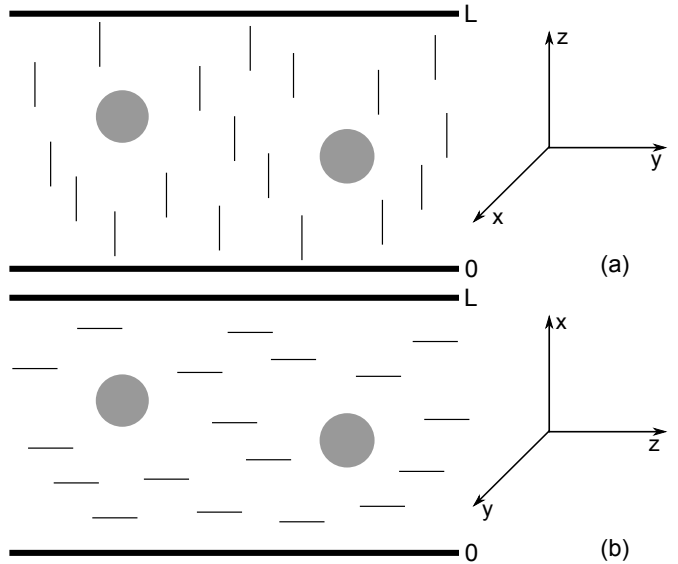}
\end{center}
\caption{(a)Homeotropic nematic cell of thickness $L$. (b) Planar nematic cell of thickness $L$}\label{CS}
\end{figure}

Here we will describe briefly the method developed in \cite{we,we2} and extend it to the case of presence of an external electric or magnetic field.
Consider axially symmetric particle of the micron or sub-micron size which may carry topological defects such as hyperbolic hedgehog, disclination ring or boojums. In the absence of the particle ground, the non-deformed state of NLC is the orientation of the director $\textbf{n}|| z, \textbf{n}=(0,0,1)$. The immersed particle induces deformations of the director in the perpendicular directions $n_{\mu}, \mu=x,y$. In close vicinity to the particle, strong deformations and topological defects arise, but beyond them, deformations become small.
In the paper \cite{lev3} this area with strong deformation and defects was called the \textit{coat}. Beyond the coat, the bulk energy of deformation may be approximately written in the form:
\begin{equation}
\tilde{F}_{bulk}=\frac{K}{2}\int  d^{3}x (\nabla n_{\mu})^{2} \label{fb3}
\end{equation}
with Euler-Lagrange equations of Laplace type:
\begin{equation}
\Delta n_{\mu}=0  \label{lap}
\end{equation}
Then director field outside the coat in the infinite LC has the form $n_{x}(\textbf{r})=p\frac{x}{r^{3}}+3c\frac{xz}{r^{5}},n_{y}(\textbf{r})=p\frac{y}{r^{3}}+3c\frac{yz}{r^{5}}$ with $p$ and $c$ being dipole and quadrupole elastic moments (we use another notation for $c$ with respect to the $\tilde{c}$ in \cite{lupe}, so that our $c=\frac{2}{3}\tilde{c}$ ).
Actually $p$ and $c$ are unknown quantities. They can be found only as asymptotics from exact solutions or from variational ansatzes. It was found in \cite{lupe} that $p=\alpha a^2$, $c=-\beta a^3$ with $a$ being the particle radius and for instance $\alpha=2.04$, $\beta=0.72$ for hyperbolic hedgehog configuration \cite{lupe}. Experiment \cite{noel} gives $\alpha=2.05$, $\beta=0.2\pm0.1$. Consider that we have found $p$ and $c$ and they \textit{are fixed}. Then we can formulate an effective theory which describes particles and interactions between them, remarkably well. 
 In order to find energy of the system: particle(s) + LC it is necessary to introduce some effective free energy functional $F_{eff}$ so that it's Euler-Lagrange equations would have the above solutions. In the \cite{lupe} it was found that in the one constant approximation with Frank constant $K$, the effective functional has the form:
\begin{equation}
F_{eff}=K\int d^{3}x\left\{\frac{(\nabla n_{\mu})^{2}}{2}-4\pi P(\textbf{x})\partial_{\mu}n_{\mu}-4\pi C(\textbf{x})\partial_{z}\partial_{\mu}n_{\mu} \right\}\label{flin}
\end{equation}
which brings Euler-Lagrange equations:
\begin{equation}
\Delta n_{\mu}=4\pi\left[\partial_{\mu}P(\textbf{x})-\partial_{z}\partial_{\mu}C(\textbf{x})\right] \label{nmu}
\end{equation}
where $P(\textbf{x})$ and $C(\textbf{x})$ are dipole and quadrupole moment densities, $\mu=x,y$ and repeated $\mu$ means summation on $x$ and $y$ like $\partial_{\mu}n_{\mu}=\partial_{x}n_{x}+\partial_{y}n_{y}$.
For the infinite space the solution has the known form: $n_{\mu}(\textbf{x})=\int d^{3}\textbf{x}' \frac{1}{\left|\textbf{x}-\textbf{x}'\right|}\left[ -\partial_{\mu}'P(\textbf{x}')+\partial_{\mu}'\partial_{z}'C(\textbf{x}') \right] $. If we consider $P(\textbf{x})=p\delta(\textbf{x})$ and $C(\textbf{x})=c\delta(\textbf{x})$ this really brings $n_{x}(\textbf{r})=p\frac{x}{r^{3}}+3c\frac{xz}{r^{5}},n_{y}(\textbf{r})=p\frac{y}{r^{3}}+3c\frac{yz}{r^{5}}$.
This means that effective functional (\ref{flin}) correctly describes the interaction between particle and LC via linear terms $-4\pi P(\textbf{x})\partial_{\mu}n_{\mu}-4\pi C(\textbf{x})\partial_{z}\partial_{\mu}n_{\mu}$. 

When the external electric or magnetic field is applied we need to add one more term to the free energy \eqref{flin} to take into account interaction between field and nematic LC. Thus, the effective free energy for the system under investigation is:
\begin{widetext}
\begin{equation}\label{F_eff}
F_{eff}^{field}=K\int d^{3}x\left\{ \frac{(\nabla n_{\mu})^2}{2}-\frac{k^2}{2}(\mathbf{en})^2-4\pi P(\mathbf{x})\partial_{\mu}n_{\mu}+4\pi \partial_{z}C(\mathbf{x})\partial_{\mu}n_{\mu} \right\}
\end{equation}
\end{widetext}
where $k^2=(4\pi K)^{-1}\Delta\varepsilon E^2$ if an electric field is applied and $k^2=K^{-1}\Delta\chi H^2$ in the case of a magnetic field; $\Delta\varepsilon=\varepsilon_{||}-\varepsilon_{\bot}$ and $\Delta\chi=\chi_{||}=\chi_{\bot}$ are dielectric and magnetic anisotropies of the NLC respectively; $\mathbf{e}||\textbf{E},\textbf{H}$ is the unit vector in the field direction.
Hereafter we will consider only electric field for convenience. But all results can be applied as well to the magnetic field case by the substitution $\frac{\Delta\varepsilon}{4\pi}E^2\to \Delta\chi H^2$. In this paper we will consider only cases when the electric field is parallel \textit{to the one of the any coordinate axis $(x,y,z)$}. When the electric field is parallel to the $z$ axis the Euler-Lagrange equations are:

\begin{equation}
\Delta n_{\mu}=k^{2}n_{\mu}+4\pi\left[\partial_{\mu}P(\textbf{x})-\partial_{z}\partial_{\mu}C(\textbf{x})\right] \label{nmuz}
\end{equation}

When the electric field is applied along one $p=x$ or $p=y$ direction, then Euler-Lagrange equations are:

\begin{equation}
\Delta n_{\mu}=-k^{2}\delta_{p\mu}n_{\mu}+4\pi\left[\partial_{\mu}P(\textbf{x})-\partial_{z}\partial_{\mu}C(\textbf{x})\right] \label{nmux}
\end{equation}
So in general case we have two different Euler-Lagrange equations for $n_{x}$ and $n_{y}$. 

In the case of confined nematic LC with the boundary conditions $n_{\mu}(\textbf{s})=0$ on the surfaces $\Sigma$ (Dirichlet boundary conditions)
the solution of EL equations has the form:
\begin{equation}
n_{\mu}(\textbf{x})=\int_{V} d^{3}\textbf{x}' G_{\mu}(\textbf{x},\textbf{x}')\left[ -\partial_{\mu}'P(\textbf{x}')+\partial_{\mu}'\partial_{z}'C(\textbf{x}') \right] \label{solmain}
\end{equation}
where $G_{\mu}$ is the Green function for $n_{\mu}$. In the case of field direction $\textbf{e}||z$  we have only one Green function $G=G_{\mu}$ which satisfies equation $(\Delta_{\textbf{x}}-k^{2})G(\textbf{x},\textbf{x}')=-4\pi \delta(\textbf{x}-\textbf{x}')$ for  $\textbf{x},\textbf{x}'\in \textbf{V}$ ($\textbf{V}$ is the volume of the bulk NLC) and $G(\textbf{x},\textbf{s})=0 $ for any \textbf{s} of the bounding surfaces $\Sigma$.

In the case of $\textbf{e}||p$, $p=x$ or $p=y$ field direction we have two different Green functions $G_{\mu}$ which satisfy equations $(\Delta_{\textbf{x}}+k^{2}\delta_{p\mu})G_{\mu}(\textbf{x},\textbf{x}')=-4\pi \delta(\textbf{x}-\textbf{x}')$ for  $\textbf{x},\textbf{x}'\in \textbf{V}$ ($\textbf{V}$ is the volume of the bulk NLC) and $G_{\mu}(\textbf{x},\textbf{s})=0 $ for any \textbf{s} of the bounding surfaces $\Sigma$.
 
 The mathematical symmetry property $G_{\mu}(\textbf{x},\textbf{x}')=G_{\mu}(\textbf{x}',\textbf{x})$ can be proved for the Green functions satisfying the Dirichtle boundary conditions by means of Green's theorem \cite{jac}.

 Consider $N$ particles in the confined NLC, so that $P(\textbf{x})=\sum_{i}p_{i}\delta(\textbf{x}-\textbf{x}_{i})$ and $C(\textbf{x})=\sum_{i}c_{i}\delta(\textbf{x}-\textbf{x}_{i})$. Then substitution (\ref{solmain}) into $F_{eff}^{field}$ brings: $F_{eff}^{field}=U^{self}+U^{interaction}$ where  $ U^{self}=\sum_{i}U_{i}^{self} $ , here $U_{i}^{self}$ is the interaction of the $i$-th particle with the bounding surfaces in the presence of the field $ U_{i}^{self}=U_{dd}^{self}+U_{dQ}^{self}+U_{QQ}^{self} $. In the general case, the interaction of the particle with bounding surfaces, (self-energy part), takes the form:  
\begin{equation}
 U_{dd}^{self}=-2\pi K p^{2}\partial_{\mu}\partial_{\mu}'H_{\mu}(\textbf{x}_{i},\textbf{x}_{i}')|_{\textbf{x}_{i}=\textbf{x}_{i}'}\label{uself}
\end{equation}
$$
 U_{dQ}^{self}=-2\pi K pc( \partial_{\mu}\partial_{\mu}'\partial_{z}'H_{\mu}(\textbf{x}_{i},\textbf{x}_{i}')+\partial_{\mu}'\partial_{\mu}\partial_{z}H_{\mu}(\textbf{x}_{i},\textbf{x}_{i}'))|_{\textbf{x}_{i}=\textbf{x}_{i}'} 
$$
 $$
 U_{QQ}^{self}=-2\pi K c^{2} \partial_{z}\partial_{z}'\partial_{\mu}\partial_{\mu}'H_{\mu}(\textbf{x}_{i},\textbf{x}_{i}')|_{\textbf{x}_{i}=\textbf{x}_{i}'}
 $$
where  $H_{\mu}(\textbf{x},\textbf{x}')=G_{\mu}(\textbf{x},\textbf{x}')-\frac{1}{|\textbf{x}-\textbf{x}'|}$ (we excluded the divergent part of self energy ).\\
Interaction energy $ U^{interaction}=\sum_{i<j}U_{ij}^{int} $.  Here $U_{ij}^{int}$ is the interaction energy between $i$ and $j$ particles: 
 $ U_{ij}^{int}= U_{dd}+U_{dQ}+U_{QQ} $:
\begin{equation}
U_{dd}=-4\pi K pp'\partial_{\mu}\partial_{\mu}'G_{\mu}(\textbf{x}_{i},\textbf{x}_{j}') \label{uint}
\end{equation}

$$
U_{dQ}=-4\pi K \left\{pc'\partial_{\mu}\partial_{\mu}'\partial_{z}'G_{\mu}(\textbf{x}_{i},\textbf{x}_{j}')+p'c\partial_{\mu}'\partial_{\mu}\partial_{z}G_{\mu}(\textbf{x}_{i},\textbf{x}_{j}')\right\}
$$

$$
U_{QQ}=-4\pi K cc'\partial_{z}\partial_{z}'\partial_{\mu}\partial_{\mu}'G_{\mu}(\textbf{x}_{i},\textbf{x}_{j}')
$$
Here unprimed quantities are used for particle $i$ and primed for particle $j$. $U_{dd}, U_{dQ}, U_{QQ}$ means dipole-dipole, dipole-quadrupole and quadrupole-quadrupole interactions, respectively. 

Formulas (\ref{uself}) and (\ref{uint}) represent general expressions for the self energy of one particle, (energy of interaction with the walls), and interparticle elastic interactions in the arbitrary confined NLC with strong anchoring conditions $n_{\mu}(\textbf{s})=0$ on the bounding surfaces and external electric or magnetic field applied parallel to the one of the coordinate axes $(x,y,z)$. 

It should be noted here that the difference between dielectric constants of the nematic host and suspended particles can give rise to an electrostatic interaction between latter \cite{lopo,dip}.
For instance dipole-dipole induced electric interaction $\textbf{F}_{electric}$ can be roughly estimated using the relation for two spheres immersed in an isotropic fluid \cite{dip}:
\begin{equation}
F_{electric}=12\pi\varepsilon_{0}\varepsilon_{LC}\left(\frac{\varepsilon-\varepsilon_{LC}}{\varepsilon+2\varepsilon_{LC}}\right)^{2}\frac{R^{6}E^{2}}{r^{4}}\label{felectr}
\end{equation}
where R is the radius of the droplets, $\varepsilon_{0}$ is the permittivity
of free space, and $\varepsilon_{LC}$ is the dielectric constant of the particle, $\varepsilon_{LC}$ is the averaged dielectric constant of the LC and $r$ is the distance between spheres. In the whole paper we do not take into account this effect and focus only on the elastic interactions $\textbf{F}_{elastic}$ between colloidal particles arising from deformations of the director field. 

\begin{figure}
\begin{center}
\includegraphics[width=\columnwidth]{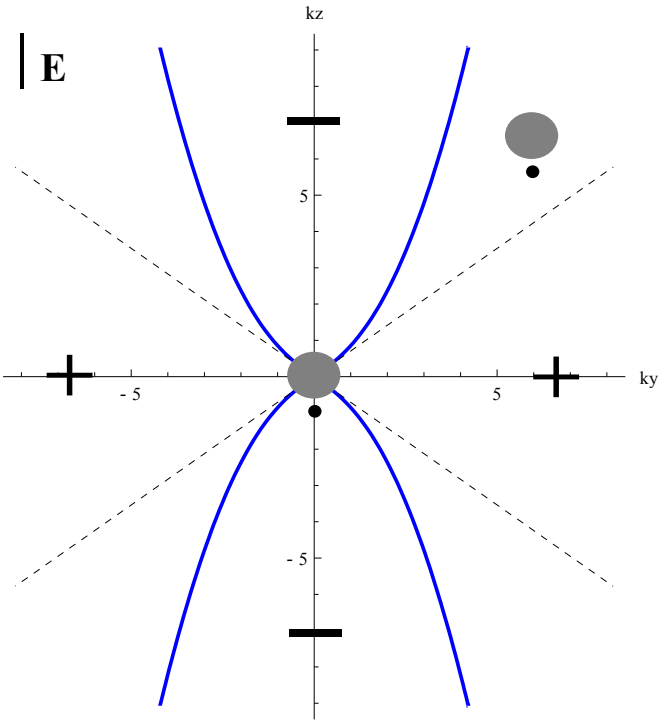}
\end{center}
\caption{(Color online) Map of the attraction and repulsion zones for dipole-dipole interaction (\ref{dd_unl}) between two particles in the infinite NLC in the presence of the electric field $\textbf{E}||z,\Delta\varepsilon>0$ as the function of the dimensionless distance $kr$, $k_{E}=E\sqrt{\Delta\varepsilon/4\pi K}$}\label{dd_unlim_p}
\end{figure}

\begin{figure}
\begin{center}
\includegraphics[width=\columnwidth]{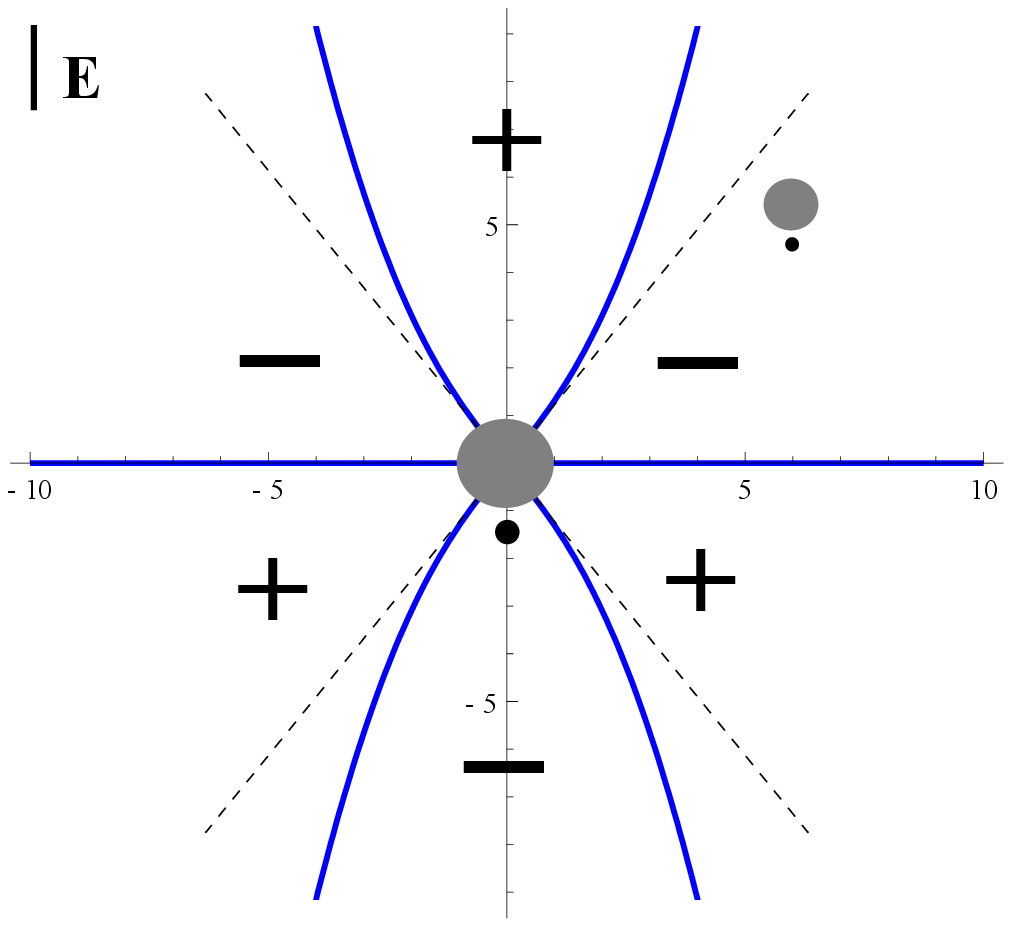}
\end{center}
\caption{(Color online) Map of the attraction and repulsion zones for dipole-quadrupole interaction (\ref{dQ_unl}) between two particles in the infinite NLC in the presence of the electric field $\textbf{E}||z,\Delta\varepsilon>0$ as the function of the dimensionless distance $kr$, $k_{E}=E\sqrt{\Delta\varepsilon/4\pi K}$}\label{dQ_unlim_p}
\end{figure}

\begin{figure}
\begin{center}
\includegraphics[width=\columnwidth]{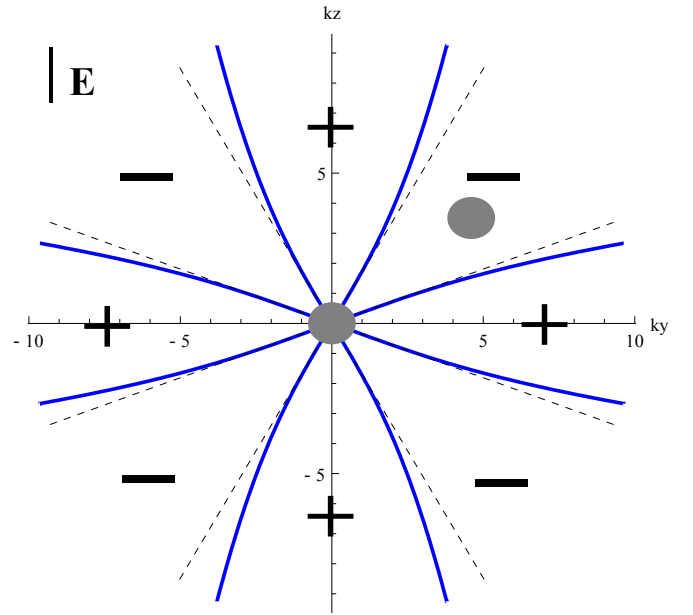}
\end{center}
\caption{(Color online) Map of the attraction and repulsion zones for dipole-dipole interaction (\ref{QQ_unl}) between two particles in the infinite NLC in the presence of the electric field $\textbf{E}||z,\Delta\varepsilon>0$ as the function of the dimensionless distance $kr$, $k_{E}=E\sqrt{\Delta\varepsilon/4\pi K}$}\label{QQ_unlim_p}
\end{figure}

 Let us consider for beginning briefly an infinite nematic liquid crystal with positive dielectric anisotropy $\Delta\varepsilon>0$. In this case the external field can be applied only along the $\mathbf{n}_{0}$, i.e. along the $z$-axis, without great change of the initial director field $\textbf{n}_{0}=(0,0,1)$ . The corresponding Euler-Lagrange equations are
\begin{equation}
\Delta n_{\mu}-k^2 n_{\mu}=4\pi [\partial_{\mu}P(\mathbf{x})-\partial_{\mu}\partial_{z}C(\mathbf{x})]
\end{equation}
where $\mu=x,y\,\,;k>0$ and $n_{\mu}(\mathbf{x}\to\infty)=0 $.
Hence the Green functions are
\begin{equation}
G_{\mu}(\mathbf{x},\mathbf{x}^{\prime})=\frac{e^{-k\left|\mathbf{x}-\mathbf{x}^{\prime} \right|}}{\left|\mathbf{x}-\mathbf{x}^{\prime} \right|}
\end{equation}  
Then using \eqref{uint} it is easy to find that all interactions between particles are exponentially screened in this case. 
\begin{widetext}
\begin{equation}
\frac{U_{\text{dd}}}{4\pi K}=\frac{pp^{\prime}}{r^{3}}\left\{ (1-3\cos^{2}\theta)(1+kr)+k^{2}r^{2}\sin^{2}\theta\right\}e^{-kr}
\label{dd_unl}
\end{equation}
\begin{equation}
\frac{U_{\text{dQ}}}{4\pi K}=(pc^{\prime}-cp^{\prime})\frac{\cos\theta}{r^{4}}\left\{ (15\cos^{2}\theta -9)(1+kr)+k^{2}r^{2}(6\cos^{2}\theta -4)-k^{3}r^{3}\sin^{2}\theta\right\}e^{-kr}
\label{dQ_unl}
\end{equation}
\begin{multline}
\frac{U_{\text{QQ}}}{4\pi K}=\frac{cc^{\prime}}{r^{5}} \{ (9-90\cos^{2}\theta +105\cos^{4}\theta)(1+kr)+k^{2}r^{2}(4-39\cos^{2}\theta +45\cos^{4}\theta)+\\
+k^{3}r^{3}(1-9\cos^{2}\theta +10\cos^{4}\theta)-k^{4}r^{4}\sin^{2}\theta\cos^{2}\theta \} e^{-kr}
\label{QQ_unl}
\end{multline}
\end{widetext}
where $r=\left|\mathbf{x}-\mathbf{x}^{\prime}\right|$ is the distance between particles and $\theta$ is the angle between $\mathbf{r}$ and $z$-axis.
Such a screening of quadrupole-quadrupole interaction was qualitatively predicted in \cite{ter1}. Maps of the attraction and repulsion zones of all interactions are presented on Fig.\ref{dd_unlim_p}, Fig.\ref{dQ_unlim_p} and Fig.\ref{QQ_unlim_p}. 

These figures show big qualitative resemblance with maps of elastic interaction \cite{we2} (see as well thin black lines on Fig.\ref{Plan_x}, Fig.\ref{Plan_y}, Fig.\ref{Plan_z} for zero field) between colloidal particles in the center of the planar nematic cell with thickness $L$. So there is an analogy between influence of the electric(magnetic) field in unlimited NLC and confinement of NLC with planar nematic cell. This resemblance gives us a possibility to estimate correspondence between influence of the electric(magnetic) field in unlimited NLC and confined NLC without external field. For instance Fig.\ref{dd_unlim_p} shows that deviation from the linear cone for dipole-dipole interaction and appearance of significant screening effects take place at distances $k_{E}^{2}r_{c}^{2}\approx2$, $r_{c}=\sqrt{2}/k_{E}$. For bigger distances screening effects appear. As well Fig.21 of \cite{we2}  (and thin black lines on Fig.\ref{Plan_x} a. as well) show that the same deviation takes place in the planar nematic cell with thickness $L$ at the distance $r_{c}^{2}/L^{2}\approx 1/2$ so that we can formulate the correspondence between influence of the electric(magnetic) field in unlimited NLC and nematic cell without field in the form:
\begin{equation}
\xi_{E}=\xi_{H}\approx L/2 \label{fc}
\end{equation}
where $\xi_{E}=1/k_{E},\xi_{H}=1/k_{H}$ are the electric(magnetic) coherence lengths. 
This means that elastic properties of the unlimited NLC in the presence of the electric(magnetic) field are qualitatively the same as of the NLC confined with nematic cell with thickness $L\approx2\xi_{E}\approx2\xi_{H}$. Physically this means that half of the cell is approximately the same as the electric(magnetic) coherence length.

H. Stark as well emphasized an analogy between confined geometries and magnetic field in \cite{stark3} and gave qualitative estimates for transition from dipole to Saturn ring configuration in nematic cell with planar anchoring observed experimentally in \cite{gu}. Using qualitative speculations he came to the estimate $\xi_{H}\approx L/2-a$ where $a$ is the radius of the particle that slightly differs from (\ref{fc}). 

In \cite{stark1} it was found that Saturn ring configuration in the presence of a magnetic field in unlimited NLC becomes stable at $a/\xi_{H}\approx0.33$.
Then using analogy (\ref{fc}) we can approximately estimate that Saturn ring configuration becomes stable in the nematic cell at the particle radius more than $a_{S}$:
\begin{equation}
a_{S}\approx0.16 L \label{as}
\end{equation}

In the experiment \cite{gu} authors reported observation of the Saturn ring configuration around glass spheres with radii 20 and 50 $\mu m$ confined in the center of planar nematic cell with thickness $L=120 \mu m$. Then (\ref{as}) gives the lower limit of the radius as $a_{S}=19 \mu m$ that allows experimental value of 20 $\mu m$ . It would be really nice to test the ratio (\ref{as}) with more accurate experimental or computer simulation test.

\begin{figure*}
\begin{center}
\includegraphics[width=\textwidth]{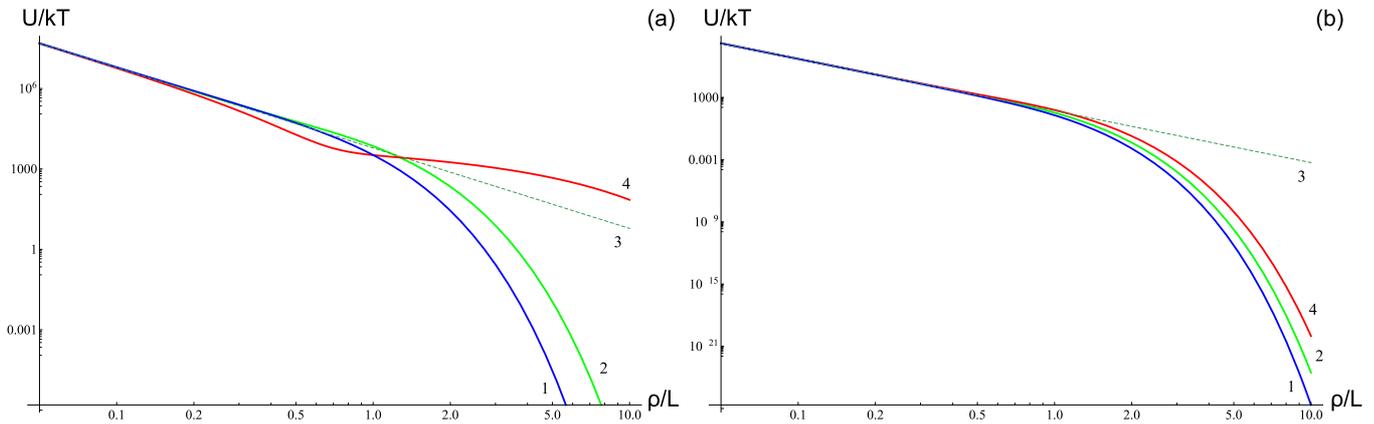}
\end{center}
\caption{ (Color online) Log-log plots of the interaction energy in $kT$ units as a function of the rescaled interparticle distance $\rho / L$. Here particle's radius $a=2.2\,\mu\text{m}$, cell thickness $L=7\,\mu\text{m}$, $K=7\,\text{pN}$, $p=p^{\prime}=2.04a^{2}$, $c=c^{\prime}=0.2a^{3}$. Dashed line 3 is the repulsion in the infinite NLC. Solid lines depict interactions in the homeotropic cell. Blue line 1 corresponds to the case $\Delta\varepsilon>0\; \text{and}\; E=0.99E_{t}$. Green line 2 corresponds to the zero-field case $E=0$. Red line 4 corresponds to $\Delta\varepsilon<0\; \text{and}\; E=0.99E_{t}$. (a) Dipole-dipole interaction. Red lines 4 -\textit{deconfinement effect}; (b) Quadrupole-quadrupole interaction. Red lines 4 - \textit{no deconfinement effect.}}\label{Hom_z}
\end{figure*}

\section{Interactions in the homeotropic cell placed in the external field}

Let us choose coordinate system with $z$-axis parallel to the undistorted director field $\mathbf{n}_{0}$ and $z=0$ at the bottom wall. The $x$- and $y$-axis lie in the cell plane (see Fig.\ref{CS}a).
There are two possible field orientations in this case. The first one is perpendicular to the cell walls, i.e. $\mathbf{e}=(0,0,1)$, and the second one is parallel to them, i.e. $\mathbf{e}=(1,0,0)$.

\subsection{Field perpendicular to the walls}\label{HomCell_z}

Here we have $\textbf{E}||z$, $(\mathbf{en})^2=n_{z}^{2}=1-\sum_{\mu}n_{\mu}^{2}$. The Euler-Lagrange equations are:
\begin{equation}\label{EL_hz}
\Delta n_{\mu}-k^2 n_{\mu}=4\pi [\partial_{\mu}P(\mathbf{x})-\partial_{\mu}\partial_{z}C(\mathbf{x})]
\end{equation} 
and $n_{\mu}(z=0)=n_{\mu}(z=L)=0$, where $\mu=x,y$.
Green functions can be easily derived from that one of the zero-field case \cite{jac}. They are:
\begin{multline}\label{G_hz}
G_{\mu}(\mathbf{x},\mathbf{x}^{\prime})=\frac{4}{L}\sum_{n=1}^{\infty}\sum_{m=-\infty}^{\infty}e^{i m (\varphi-\varphi^{\prime})}\sin\frac{n \pi z}{L}\times\\
\times\sin\frac{n \pi z^{\prime}}{L} I_{m}(\lambda_{n} \rho_{<})K_{m}(\lambda_{n}\rho_{>})
\end{multline}
where $I_{m}$, $K_{m}$ are modified Bessel functions, $\tan\varphi=\frac{y}{x}$, $\tan\varphi^{\prime}=\frac{y^{\prime}}{x^{\prime}}$, $\lambda_{n}=\sqrt{\frac{n^2 \pi^2}{L^2}+k^2}$. $\rho_{<}$ is the smaller of the  $\rho=\sqrt{x^2+y^2}$ and $\rho^{\prime}=\sqrt{x^{\prime 2}+y^{\prime 2}}$. Then \eqref{uint} brings dipole-dipole interaction in the homeotropic cell:
\begin{equation}\label{U_dd_hz}
\frac{U_{\text{dd}}}{16\pi K }=\frac{p p^{\prime}}{L}\sum_{n=1}^{\infty}\lambda_{n}^{2}\sin\frac{n \pi z}{L}\sin\frac{n \pi z^{\prime}}{L}K_{0}(\lambda_{n}\rho)
\end{equation}
here $\rho=\sqrt{(y-y^{\prime})^{2}+(x-x^{\prime})^{2}}$ is the horizontal projection of the distance between particles. It is easy to find that particles with $z=z^{\prime}$ repel when their dipole moments are parallel $pp^{\prime}>0$, and attract when $pp^{\prime}<0$.  

Similar dipole-quadrupole interaction is:
\begin{equation}\label{U_dQ_hz}
\begin{gathered}
\frac{U_{\text{dQ}}}{16\pi K }=\frac{1}{L^2}\sum_{n=1}^{\infty}\lambda_{n}^{2} n \pi K_{0}(\lambda_{n}\rho)\times\\
\times \left[ p c^{\prime}\sin\frac{n \pi z}{L}\cos\frac{n \pi z^{\prime}}{L}+c p^{\prime}\cos\frac{n \pi z}{L}\sin\frac{n \pi z^{\prime}}{L}\right]
\end{gathered}
\end{equation}
And quadrupole-quadrupole interaction takes the form:
\begin{equation}\label{U_QQ_hz}
\frac{U_{\text{QQ}}}{16\pi K }=\frac{c c^{\prime}}{L^3}\sum_{n=1}^{\infty}\lambda_{n}^{2}n^2 \pi^2 \cos\frac{n \pi z}{L}\cos\frac{n \pi z^{\prime}}{L}K_{0}(\lambda_{n}\rho)
\end{equation}
When product $cc^{\prime}$ is positive quadrupole-quadrupole interaction between particles with the same $z\text{-coordinates}$ is repulsive, when $cc^{\prime}<0$ interaction it is attractive. 

Application of \eqref{U_dd_hz} for two spherical particles with $z=z^{\prime}=\frac{L}{2}$ is shown in Fig.\ref{Hom_z}a. In the limit of small distances $\rho \ll L$ interaction energy tends to the corresponding value in the unbounded NLC regardless of the field strength and nematic's anisotropy: $U_{\text{dd}}\to\frac{4\pi K pp^{\prime}}{\rho^{3}}$. If the cell is in the zero-field (see Fig.\ref{Hom_z}a, line 2) then for distances $\rho >1.2L$ exponential decay of interaction takes place \cite{we}. 

Let us take a look at $\lambda_{n}=\sqrt{\frac{n^2 \pi^2}{L^2}+k^2}$. For further analysis it is convenient to rewrite it as
\begin{equation}
\lambda_{n}=\frac{n \pi}{L}\sqrt{1+\frac{\text{sgn}(\Delta\varepsilon)}{n^2}\left(\frac{E}{E_{t}}\right)^2}
\end{equation}
where $E_{t}$ is the Fredericks threshold electric field, $E_{t}=\frac{\pi}{L}\sqrt{\frac{4\pi K}{|\Delta\varepsilon|}}$. One can compare formulas \eqref{U_dd_hz}-\eqref{U_QQ_hz} with results of \cite{we} and find that an external field gives rise to a set of the effective cell thicknesses
\begin{equation}\label{L_eff}
L_{n}^{\text{eff}}=L \sqrt{\frac{n^2 E_{t}^2}{n^2 E_{t}^2+\text{sgn}(\Delta\varepsilon)E^2}}
\end{equation}
and thereby affects the interaction between particles. For zero field $E=0$ we have $L_{n}^{\text{eff}}=L$. It is clearly seen that when $\Delta\varepsilon>0$ the external electric field decreases effective thickness  $L_{n}^{\text{eff}}\Rightarrow 0$ ($E\Rightarrow\infty$) of the cell $L$ and increases screening strength (see Fig.\ref{Hom_z}a, line 1).

  A completely different picture we can see in the nematic with negative anisotropy $\Delta\varepsilon<0$. In this case the maximum value of the applied electric field is the Fredericks threshold value $E=E_{t}$, after which Fredericks transition takes place and initial ground state $\textbf{n}_{0}=(0,0,1)$ becomes broken. So allowed values of the electric field are $0\leq E <E_{t}$ (remember that we are bounded with the one constant approximation $K_{1}=K_{2}=K_{3}=K$ and strong anchoring on the cell planes). So gradual increasing of the field strength below $E_{t}$ makes $L_{1}^{\text{eff}}\Rightarrow \infty$ that is the \textit{Deconfinement effect} for dipole-dipole interaction between dipole particles in the center of the nematic cell near the Fredericks transition point(see Fig.\ref{Hom_z}a, line 4). Here at the distances $0.7L<\rho<8L$ energy of dipole-dipole interaction decreases even more slowly than in the infinite NLC. 
Actually this means that influence of the bounding surface planes disappears near the critical point $E_{t}$ for dipole-dipole interaction. Actually all $L_{n}^{\text{eff}}$ for $n\geq 2$ remain finite. It is interesting that this Deconfinement effect \textit{is absent for quadrupole-quadrupole interaction} between particles in the center of the cell (see Fig.\ref{Hom_z}b, line 4). Actually (\ref{U_QQ_hz}) shows that the first nonzero term in this case is $n=2$ , not $n=1$, and all $L_{n}^{\text{eff}}$ for $n\geq 2$ remain finite. So approaching the Fredericks transition only weakens the screening effect a little bit (Fig.\ref{Hom_z}b).

\subsection{Field parallel to the walls}\label{HomCell_x}

\begin{figure*}
\begin{center}
\includegraphics[width=\textwidth]{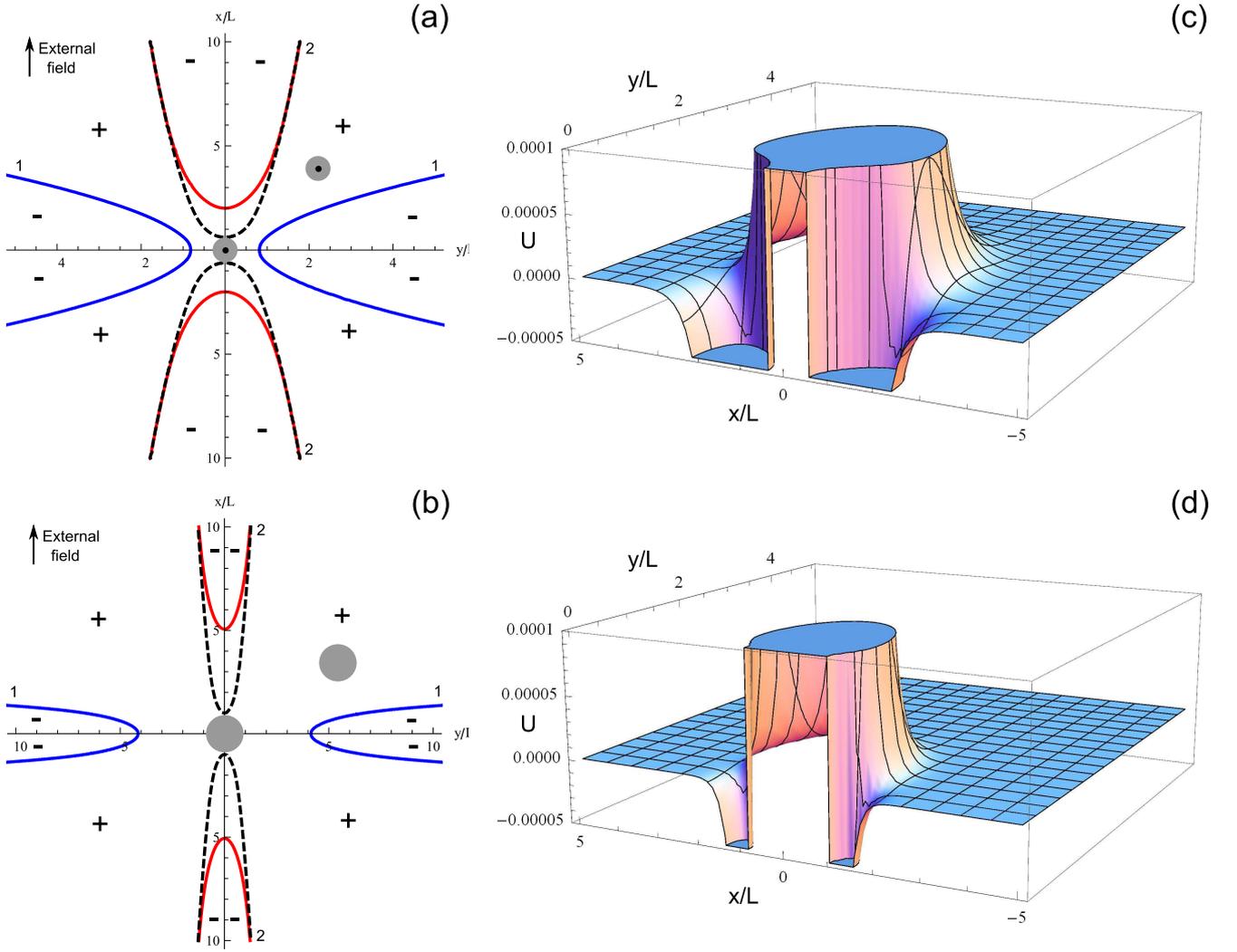}
\end{center}
\caption{ (Color online) Map of the attraction and repulsion zones of interaction between two particles in the middle of the homeotropic cell under external field $E=0.99E_{t}$. Field is parallel to the cell walls $\textbf{E}||x$. Blue lines 1 correspond to the positive $\Delta\varepsilon$, red lines 2 correspond to the negative $\Delta\varepsilon$. Dashed lines depict zones in $E=2E_{t}$ and $\Delta\varepsilon<0$. Sign ``-'' means attraction, ``+'' means repulsion. (a) Dipole-dipole interaction, $pp^{\prime}>0$. (b) Quadrupole-quadrupole interaction, $cc^{\prime}>0$. (c) Dimensionless energy of dipole-dipole interaction. (d) Dimensionless energy of quadrupole-quadrupole interaction. $E=2E_{t}$ and $\Delta\varepsilon<0$ in both graphs. }\label{Hom_x}
\end{figure*}

\begin{figure}
\begin{center}
\includegraphics[width=\columnwidth]{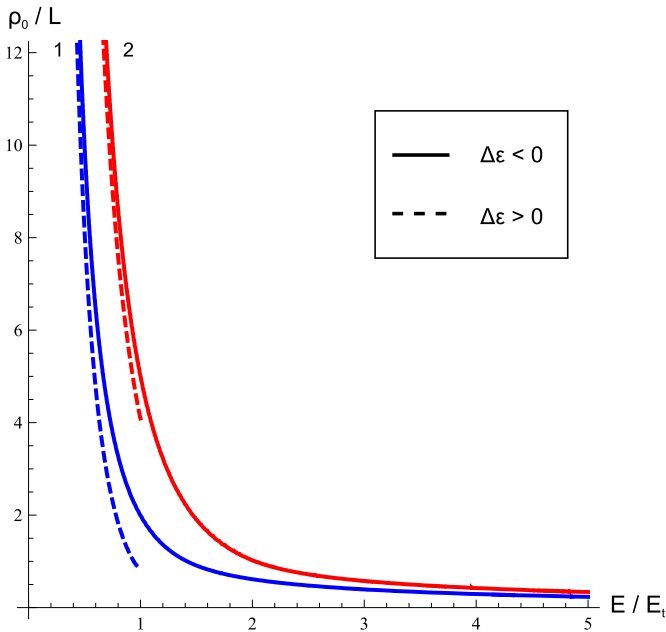}
\end{center}
\caption{ (Color online) Equilibrium distance between dipoles (blue lines~1) and quadrupoles (red lines~2) as a function of the field strength.}\label{rho}
\end{figure}

Here we consider $\textbf{E}||x$, $(\mathbf{en})^2=n_{x}^{2}$. So that we have different Euler-Lagrange equations for $n_{x}$ and $n_{y}$:
\begin{equation}\label{EL_hx}
\begin{gathered}
\Delta n_{x}+k^2 n_{x}=4\pi [\partial_{x}P(\mathbf{x})-\partial_{x}\partial_{z}C(\mathbf{x})]\\
\Delta n_{y}=4\pi [\partial_{y}P(\mathbf{x})-\partial_{y}\partial_{z}C(\mathbf{x})]
\end{gathered}
\end{equation}
with boundary conditions $n_{\mu}(z=0)=n_{\mu}(z=L)=0$, $\mu=x,y$. To get the corresponding Green functions one can use \eqref{G_hz} with substitutions $k^2 \to -k^2$ and $ k^2=0 $ for $G_{x}$ and $G_{y}$, respectively. Then \eqref{uint} brings dipole-dipole interaction:
\begin{equation}\label{U_dd_hx}
\frac{U_{\text{dd}}}{8\pi K}=\frac{p p^{\prime}}{L}\sum_{n=1}^{\infty}\sin\frac{n\pi z}{L} \sin\frac{n\pi z^{\prime}}{L} \left( A_{n}+B_{n}\cos 2\theta \right)
\end{equation}
where $\theta$ is the azimuthal angle between $\rho$ and $x$.
\begin{equation*}
\begin{gathered}
A_{n}=\lambda_{n}^{2}K_{0}(\lambda_{n}\rho)+\mu_{n}^{2}K_{0}(\mu_{n}\rho)\\
B_{n}=\lambda_{n}^{2}K_{2}(\lambda_{n}\rho)-\mu_{n}^{2}K_{2}(\mu_{n}\rho)
\end{gathered}
\end{equation*}
with $\mu=\frac{n\pi}{L}$ and $\lambda_{n}=\sqrt{\frac{n^{2}\pi^{2}}{L^{2}}-k^2}$. Dipole-quadrupole interaction is:
\begin{equation}\label{Q_dQ_hx}
\begin{gathered}
\frac{U_{\text{dQ}}}{8\pi K}=\frac{1}{L^2}\sum_{n=1}^{\infty}n\pi \left(A_{n}+B_{n}\cos 2\theta \right)\times\\
\times\left( pc^{\prime}\sin\frac{n\pi z}{L} \cos\frac{n\pi z^{\prime}}{L}+p^{\prime}c\cos\frac{n\pi z}{L} \sin\frac{n\pi z^{\prime}}{L}\right)
\end{gathered}
\end{equation}
And quadrupole-quadrupole interaction is:
\begin{equation}\label{U_QQ_hx}
\frac{U_{\text{QQ}}}{8\pi K}=\frac{c c^{\prime}}{L^{3}}\sum_{n=1}^{\infty}n^{2}\pi^{2}\cos\frac{n\pi z}{L} \cos\frac{n\pi z^{\prime}}{L} \left( A_{n}+B_{n}\cos 2\theta \right)
\end{equation}
In previous Sec.\ref{HomCell_z} we have found that interaction between particles with $z=z^{\prime}=\frac{L}{2}$ and positive products $pp^{\prime}>0$, $cc^{\prime}>0$ is isotropic and repulsive throughout the cell plane. Expressions \eqref{U_dd_hx} and \eqref{U_QQ_hx} show that external field, when it is applied parallel to the walls, breaks the symmetry of interaction and induces \textit{zones of attraction} (see Fig.\ref{Hom_x}a and \ref{Hom_x}b). Energy has a minimum at the points $\rho_{0}$ of intersection between zones borders and coordinate axes (see Fig.\ref{Hom_x}c and \ref{Hom_x}d). Hence the distance $\rho_{0}$ between the center and intersection points  is the equilibrium distance between particles. It depends on the field strength: the stronger the field, the smaller the $\rho_{0}$. Figure~\ref{rho} shows that the particles suspended in a nematic with negative anisotropy can be can be arbitrary close. When $\Delta\varepsilon>0$ the field strength must be less than $E_{t}$ to avoid the Fredericks transition, therefore, for dipoles we have $\rho_{0} > 0.8\,L$ and for quadrupoles  $\rho_{0} > 4.1\,L$. This \textit{effect of attraction and stabilization} of the colloidal particles is new and have never been reported before. 

\begin{figure*}
\begin{center}
\includegraphics[width=\textwidth]{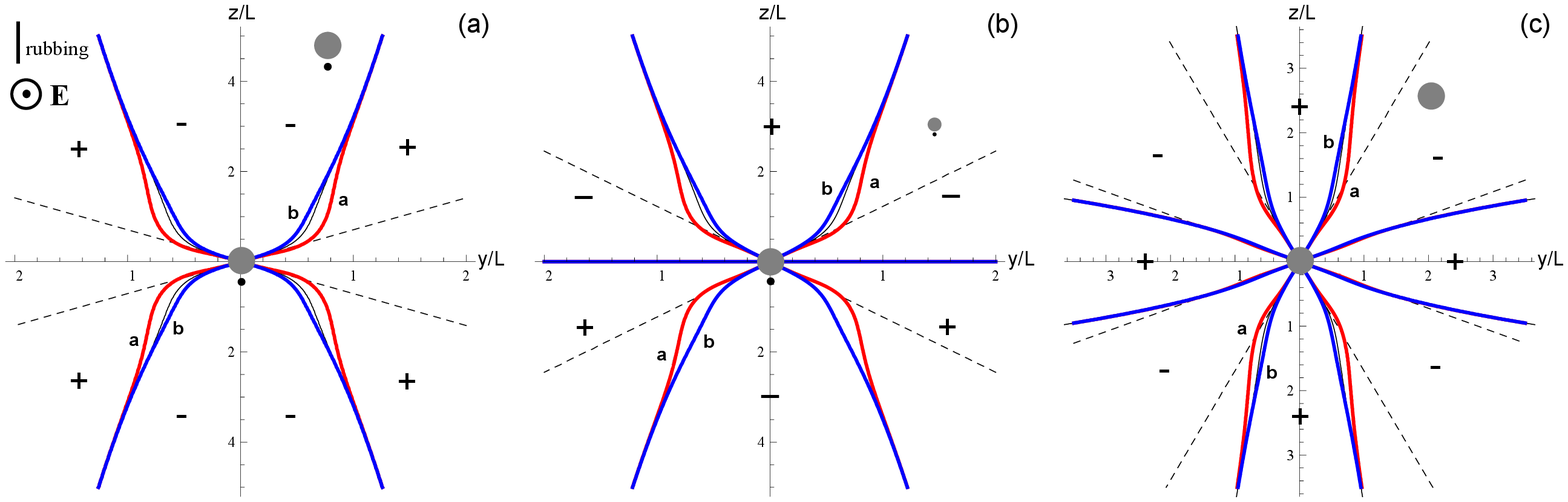}
\end{center}
\caption{(Color online) Map of the attraction and repulsion zones of interaction between two particles in the middle of the planar cell with rubbing along $z$ and thickness $L$  placed in the perpendicular field $\textbf{E}||x$, $E=0.99E_{t}$. Dashed lines are zone's borders in the unlimited NLC. Thick red lines $a$ correspond to the positive $\Delta\varepsilon$, thick blue lines $b$ correspond to the negative $\Delta\varepsilon$. Thin black lines are zone's borders in the zero-field. Sign ``-'' means attraction, ``+'' means repulsion. (a) Dipole-dipole interaction, $p p^{\prime}>0$. (b) Dipole-quadrupole interaction. The larger particle is in the center, $pc^{\prime}>cp^{\prime}$. (c) Quadrupole-quadrupole interaction, $cc^{\prime}>0$.}\label{Plan_x}
\end{figure*}

\section{Interactions in the planar cell placed in an external field}

Coordinate system is depicted in Fig.\ref{CS}b. Here we have three possible field orientations. The first one is perpendicular to the cell walls, i.e. $\mathbf{e}=(1,0,0)$. The second one is parallel to the walls and perpendicular to $\mathbf{n}_{0}$, i.e. $\mathbf{e}=(0,1,0)$. And finally, external field may be parallel to the ground director state $\mathbf{n}_{0}$, i.e. $\mathbf{e}=(0,0,1)$.

\subsection{Field perpendicular to the cell planes}

Here we consider $\textbf{E}||x$, $(\mathbf{en})^{2}=n_{x}^{2}$. The Euler-Lagrange equations are the same as \eqref{EL_hx}. But $n_{\mu}$ have another boundary conditions ($z$ is replaced with $x$), so that $n_{\mu}(x=0)=n_{\mu}(x=L)=0$, $\mu=x,y$ . Hence, the Green functions can be constructed from 
\eqref{G_hz} replacing $(x,y,z)\rightarrow(y,z,x)$ as well as $k^{2}\rightarrow-k^{2}$ and $k^{2}=0$ for $x$ and $y$ respectively, so that:
 
\begin{multline*}
G_{x}(\mathbf{x},\mathbf{x}^{\prime})=\frac{4}{L}\sum_{n=1}^{\infty}\sum_{m=-\infty}^{\infty}e^{i m (\varphi-\varphi^{\prime})}\sin\frac{n \pi x}{L}\times\\
\times\sin\frac{n \pi x^{\prime}}{L} I_{m}(\lambda_{n} \rho_{<})K_{m}(\lambda_{n}\rho_{>})
\end{multline*}
\begin{multline}\label{G_px}
G_{y}(\mathbf{x},\mathbf{x}^{\prime})=\frac{4}{L}\sum_{n=1}^{\infty}\sum_{m=-\infty}^{\infty}e^{i m (\varphi-\varphi^{\prime})}\sin\frac{n \pi x}{L}\times\\
\times\sin\frac{n \pi x^{\prime}}{L} I_{m}(\mu_{n} \rho_{<})K_{m}(\mu_{n}\rho_{>})
\end{multline}
where $\lambda_{n}=\sqrt{\frac{n^2 \pi^2}{L^2}-k^2}$, $\mu_{n}=\frac{n\pi}{L}$. $\rho_{<}$ is the smaller of the  $\rho=\sqrt{z^2+y^2}$ and $\rho^{\prime}=\sqrt{z^{\prime 2}+y^{\prime 2}}$, $\tan\varphi=\frac{y}{z}$, $\tan\varphi^{\prime}=\frac{y^{\prime}}{z^{\prime}}$. Then expressions \eqref{uint} give us all desired interactions in the planar cell. Dipole-dipole elastic interaction is given by:
\begin{equation}\label{U_dd_plan}
U_{\text{dd}}=\frac{16\pi K p p^{\prime}}{L^3}\left( F_{1}-F_{2}\cos^{2}\theta \right)
\end{equation}
with $F_{1}$ and $F_{2}$ being:
\begin{multline*}
F_{1}=\sum_{n=1}^{\infty} \frac{L^{2}\mu_{n}^{2}}{2}\sin\frac{n\pi x}{L}\sin\frac{n\pi x^{\prime}}{L}\left[K_{0}(\mu_{n}\rho)+K_{2}(\mu_{n}\rho) \right]-\\
-n^{2}\pi^{2}\cos\frac{n\pi x}{L}\cos\frac{n\pi x^{\prime}}{L}K_{0}(\lambda_{n}\rho)
\end{multline*}
\begin{equation}\label{FF}
F_{2}=\sum_{n=1}^{\infty}L^{2}\mu_{n}^{2}\sin\frac{n\pi x}{L}\sin\frac{n\pi x^{\prime}}{L}K_{2}(\mu_{n}\rho)
\end{equation}
where $\rho=\sqrt{(y-y^{\prime})^{2}+(z-z^{\prime})^{2}}$ is the horizontal projection of the distance between particles and $\theta$ is the azimuthal angle between $\rho$ and $z$.

Dipole-quadrupole interaction takes the form:
\begin{equation}\label{U_dQ_plan}
U_{\text{dQ}}=\frac{16\pi K}{L^{4}}\left(p c^{\prime}-c p^{\prime} \right)\cos\theta\left(C_{1}+C_{2}\cos^{2}\theta \right)
\end{equation}
where
\begin{equation*}
\begin{gathered}
C_{1}=L\left( F_{1\rho}^{\prime}-\frac{2 F_{2}}{\rho}\right)\\
C_{2}=L\left( \frac{2 F_{2}}{\rho}-F_{2\rho}^{\prime}\right)
\end{gathered}
\end{equation*}
Quadrupole-quadrupole interaction is:
\begin{equation}\label{U_QQ_plan}
U_{\text{QQ}}=\frac{16\pi K c c^{\prime}}{L^5}\left( D_{1}+D_{2}\cos^{2}\theta +D_{3}\cos^{4}\theta \right)
\end{equation}
where
\begin{equation*}
\begin{gathered}
D_{1}=L^{2}\left( \frac{2 F_{2}}{\rho^{2}} - \frac{F_{1\rho}^{\prime}}{\rho}\right)\\
D_{2}=L^{2}\left( -\frac{10 F_{2}}{\rho^{2}} + \frac{5 F_{2\rho}^{\prime}}{\rho}+\frac{F_{1\rho}^{\prime}}{\rho}-F_{1\rho\rho}^{\prime\prime}\right)\\
D_{3}=L^{2}\left( \frac{8 F_{2}}{\rho^{2}}-\frac{5 F_{2\rho}^{\prime}}{\rho}+F_{2\rho\rho}^{\prime\prime}\right)
\end{gathered}
\end{equation*}

Maps of the attraction and repulsion zones for interaction  between particles with $x=x^{\prime}=\frac{L}{2}$ are presented in Fig.\ref{Plan_x}. In the limit of small distances $\rho\ll L$ all interactions are the same as in the unbounded nematic: $U_{\text{dd}}\to\frac{4\pi K pp^{\prime}}{\rho^{3}}(1-3\cos^{2}\theta)$, $U_{\text{dQ}}\to\frac{4\pi K}{\rho^{4}}(pc^{\prime}-cp^{\prime})(15\cos^{2}\theta-9)$, $U_{\text{QQ}}\to\frac{4\pi K cc^{\prime}}{\rho^{5}}(9-90\cos^{2}\theta+105\cos^4\theta)$. When $\rho>10L$ zone's borders coincide with those in the cell under zero-field. At the intermediate distances scale field moves zone's borders closer to those in the unbounded NLC when $\Delta\varepsilon>0$ (see Fig.\ref{Plan_x}, lines $a$). If the nematic has a negative $\Delta\varepsilon$ then field shrinks parabola-like zones and, in fact, enhances effects caused by the bounding walls (see Fig.\ref{Plan_x}, lines $b$).

\subsection{Field parallel to the cell planes and perpendicular to the rubbing direction}

\begin{figure*}
\begin{center}
\includegraphics[width=\textwidth]{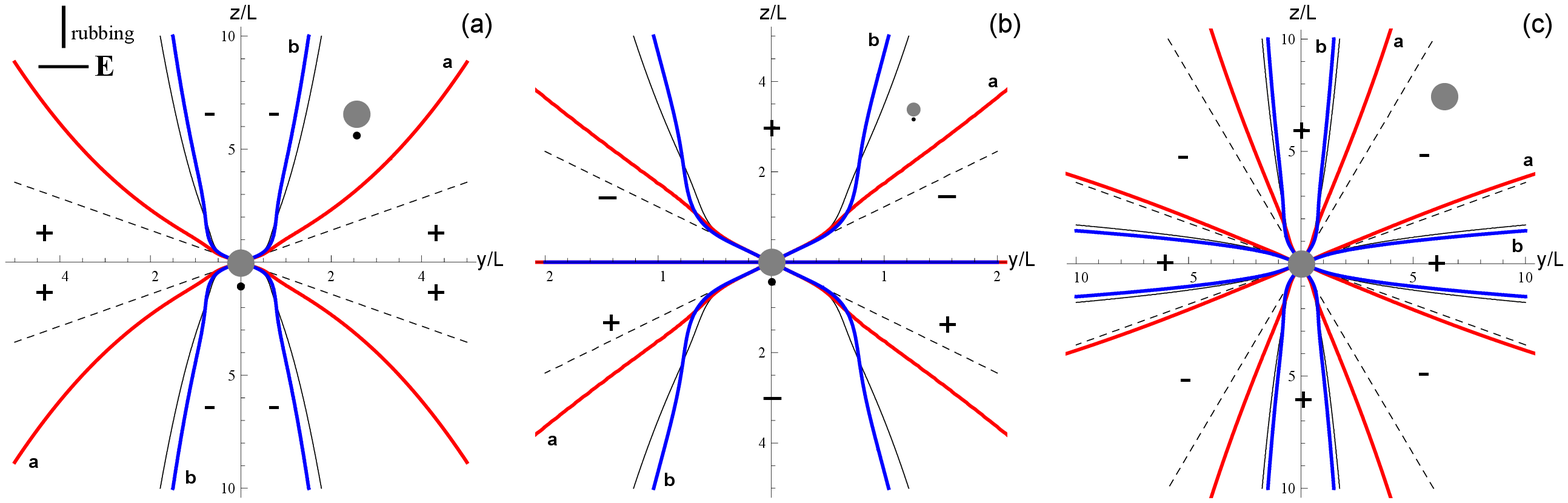}
\end{center}
\caption{(Color online) Map of the attraction and repulsion zones of interaction between two particles in the middle of the planar cell with rubbing along $z$ and thickness $L$  placed in the electric field applied along the $y\text{-axis}, \textbf{E}||y$ (see Fig.\ref{CS}b), $E=0.99E_{t}$. Dashed lines are zone's borders in the unlimited NLC. Thick red lines $a$ correspond to the positive $\Delta\varepsilon$, thick blue lines $b$ correspond to the negative $\Delta\varepsilon$. Thin black lines are zone's borders in the zero-field. Sign ``-'' means attraction, ``+'' means repulsion. (a) Dipole-dipole interaction, $pp^{\prime}>0$. (b) Dipole-quadrupole interaction. The larger particle is in the center, $pc^{\prime}>cp^{\prime}$. (c) Quadrupole-quadrupole interaction, $cc^{\prime}>0$.}\label{Plan_y}
\end{figure*}

\begin{figure}
\includegraphics[clip=,width=\linewidth]{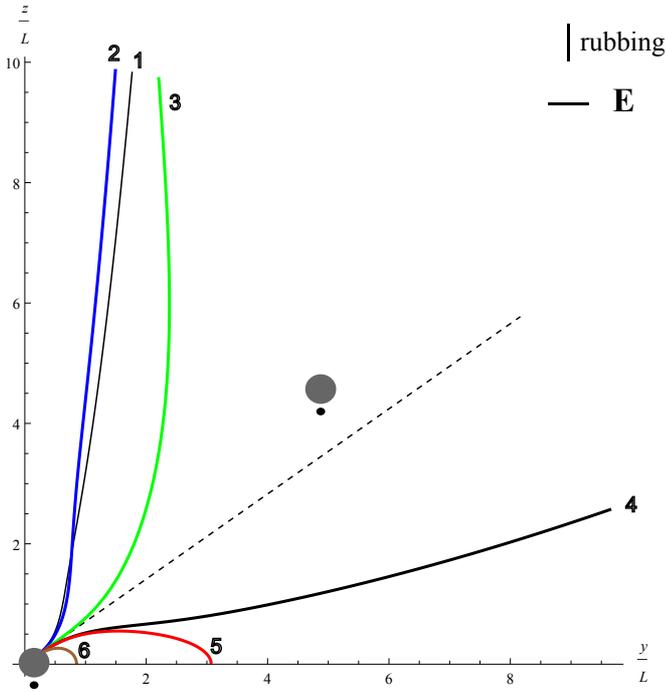}
\caption{(Color online) Dynamics of the attraction and repulsion zones of dipole-dipole interaction between two particles in the middle of the planar cell with $\Delta\varepsilon<0$ in the electric field applied along the $y\text{-axis}, \textbf{E}||y$ with increasing of the field strength $E$. The upper right quarter of the map is depicted. Dashed line is the zone's border in the unlimited NLC. Thin black line 1 corresponds to the $E=0$, thick blue line 2 - $E=0.99E_{t}$, thick green line 3 - $E=1.6E_{t}$, thick black line 4 - $E=1.73E_{t}$, thick red line 5 - $E=1.74E_{t}$, thick brown line 6 - $E=2E_{t}$. Attraction zone is above each line, repulsion zone is below each line, $pp^{\prime}>0$ (see Fig.\ref{dfabove}, Fig.\ref{dfbelow} as well). Collapse of the lateral tales into dumbbell-shaped region occurs at $E=1.74E_{t}$.}\label{ddmany}
\end{figure}

\begin{figure}
\includegraphics[clip=,width=\linewidth]{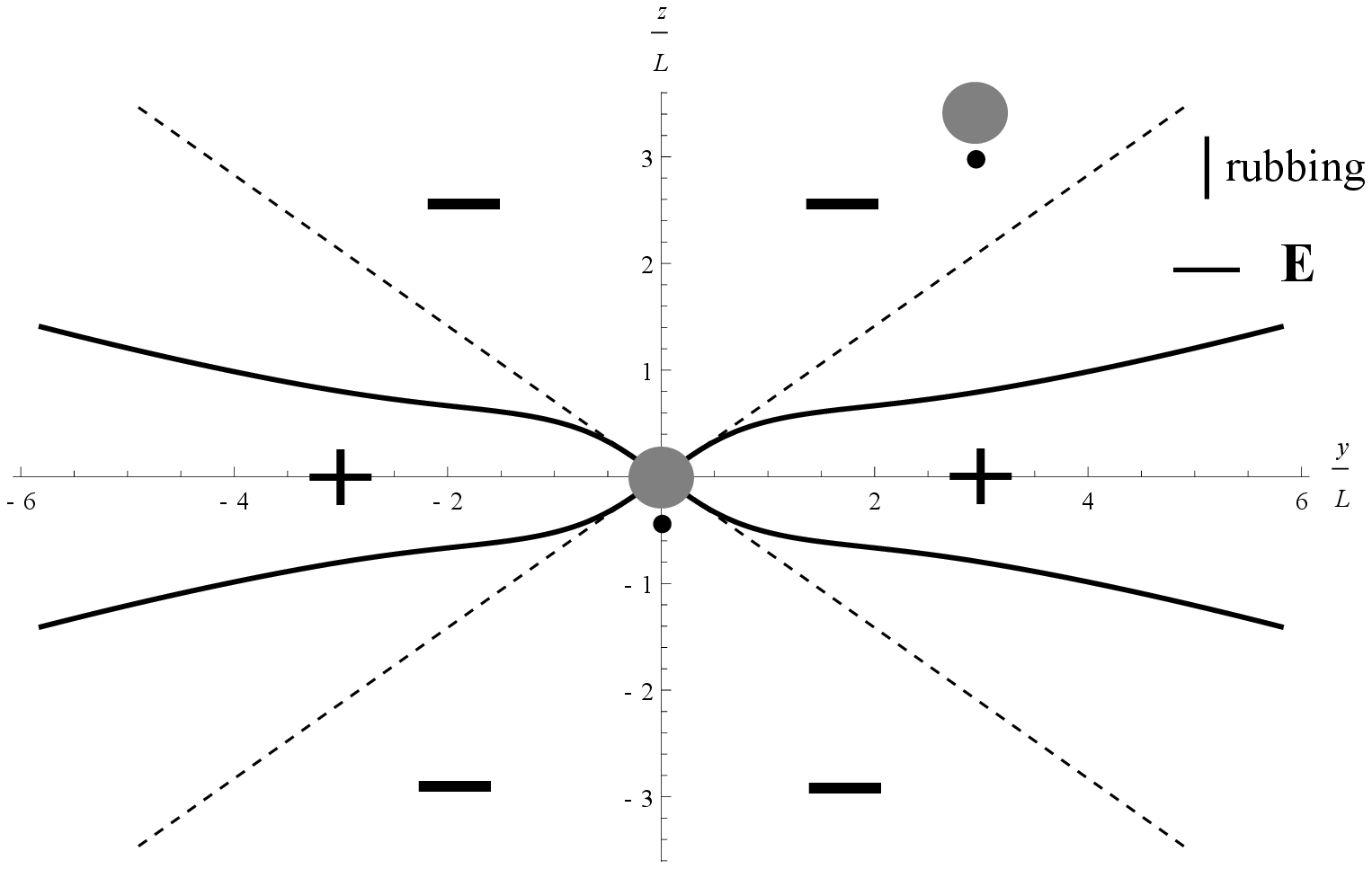}
\caption{Map of the attraction and repulsion zones of dipole-dipole interaction between two particles in the middle of the planar cell with $\Delta\varepsilon<0$ placed in the electric field applied along the $y\text{-axis}, \textbf{E}||y$, $E=1.73E_{t}$. Dashed lines are zone's borders in the unlimited NLC.  Sign ``-'' means attraction, ``+'' means repulsion, $pp^{\prime}>0$.}\label{dfabove}
\end{figure}

\begin{figure}
\includegraphics[clip=,width=\linewidth]{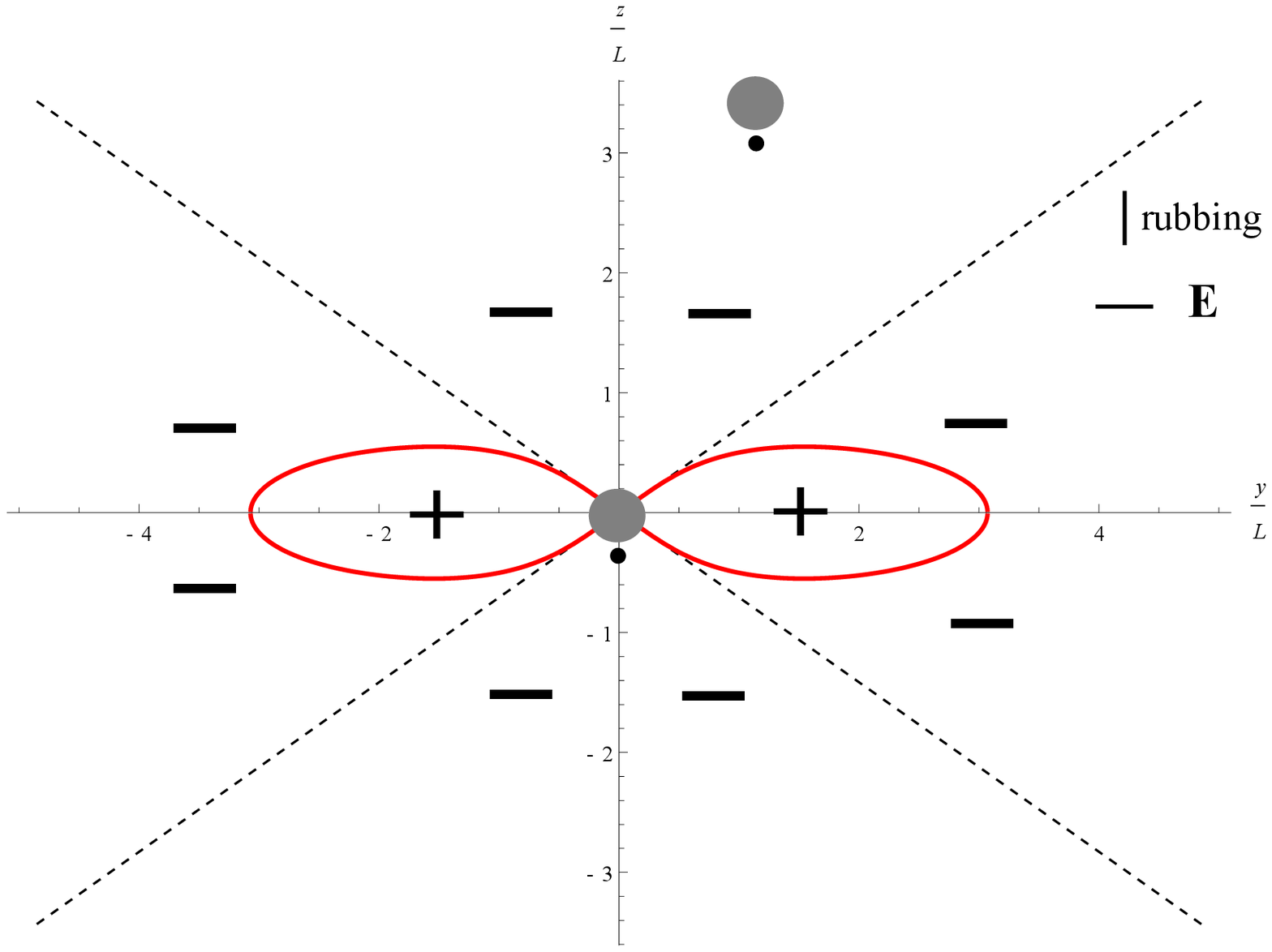}
\caption{(Color online) Map of the attraction and repulsion zones of dipole-dipole interaction between two particles in the middle of the planar cell with $\Delta\varepsilon<0$ placed in the electric field applied along the $y\text{-axis}, \textbf{E}||y$, $E=1.74E_{t}$. Dashed lines are zone's borders in the unlimited NLC. Collapse of the lateral tales into dumbbell-shaped region occurs at $E_{col}=1.74E_{t}$ (see Fig.\ref{dfbelow}). Sign ``-'' means attraction, ``+'' means repulsion, $pp^{\prime}>0$.}\label{dfbelow}
\end{figure}

\begin{figure}
\includegraphics[clip=,width=\linewidth]{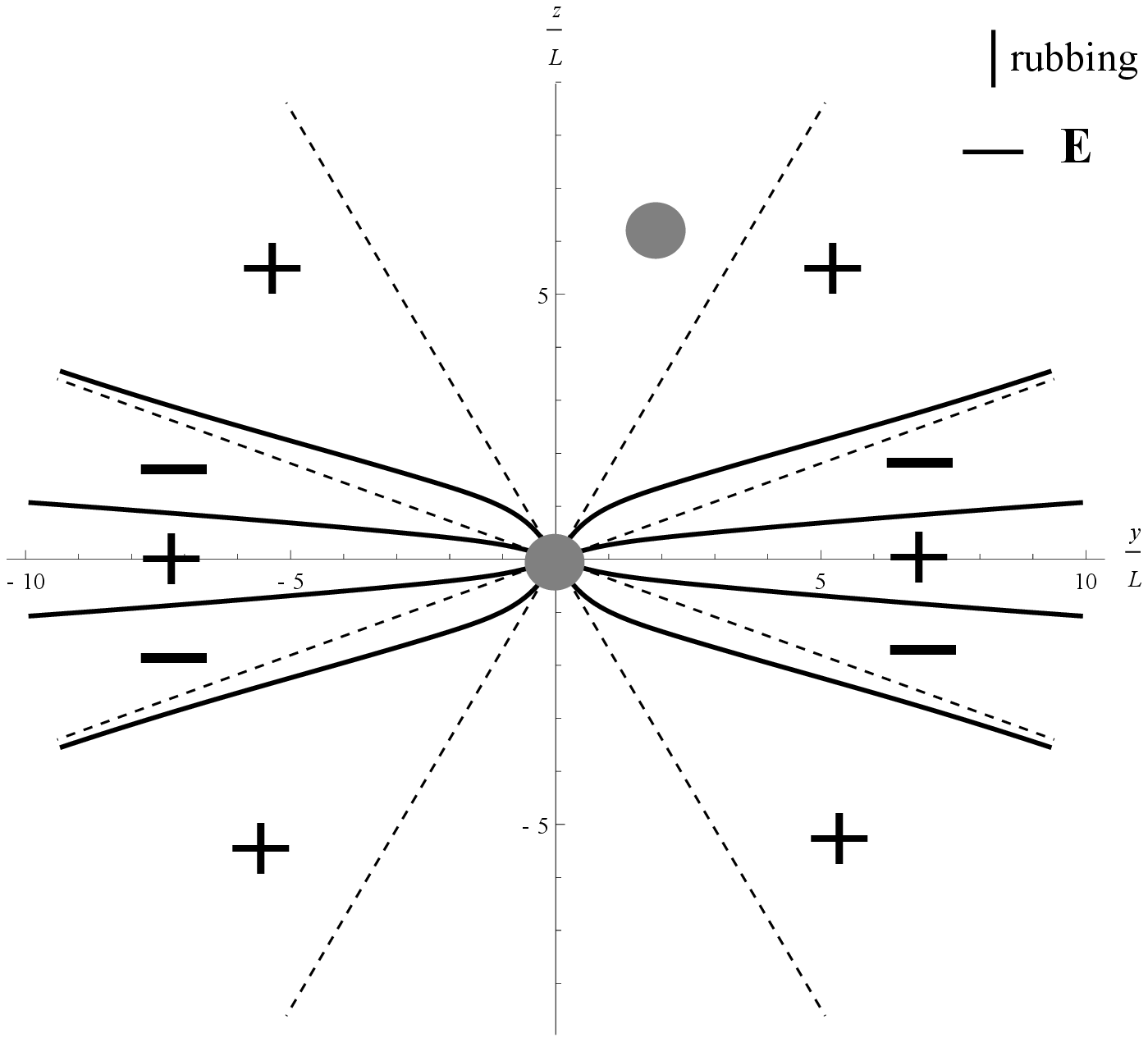}
\caption{Map of the attraction and repulsion zones of quadrupole-quadrupole interaction between two particles in the middle of the planar cell with $\Delta\varepsilon<0$ placed in the electric field applied along the $y\text{-axis}, \textbf{E}||y$, $E=1.73E_{t}$. Dashed lines are zone's borders in the unlimited NLC.  Sign ``-'' means attraction, ``+'' means repulsion, $cc^{\prime}>0$.}\label{qqbelow}
\end{figure}

\begin{figure}
\includegraphics[clip=,width=\linewidth]{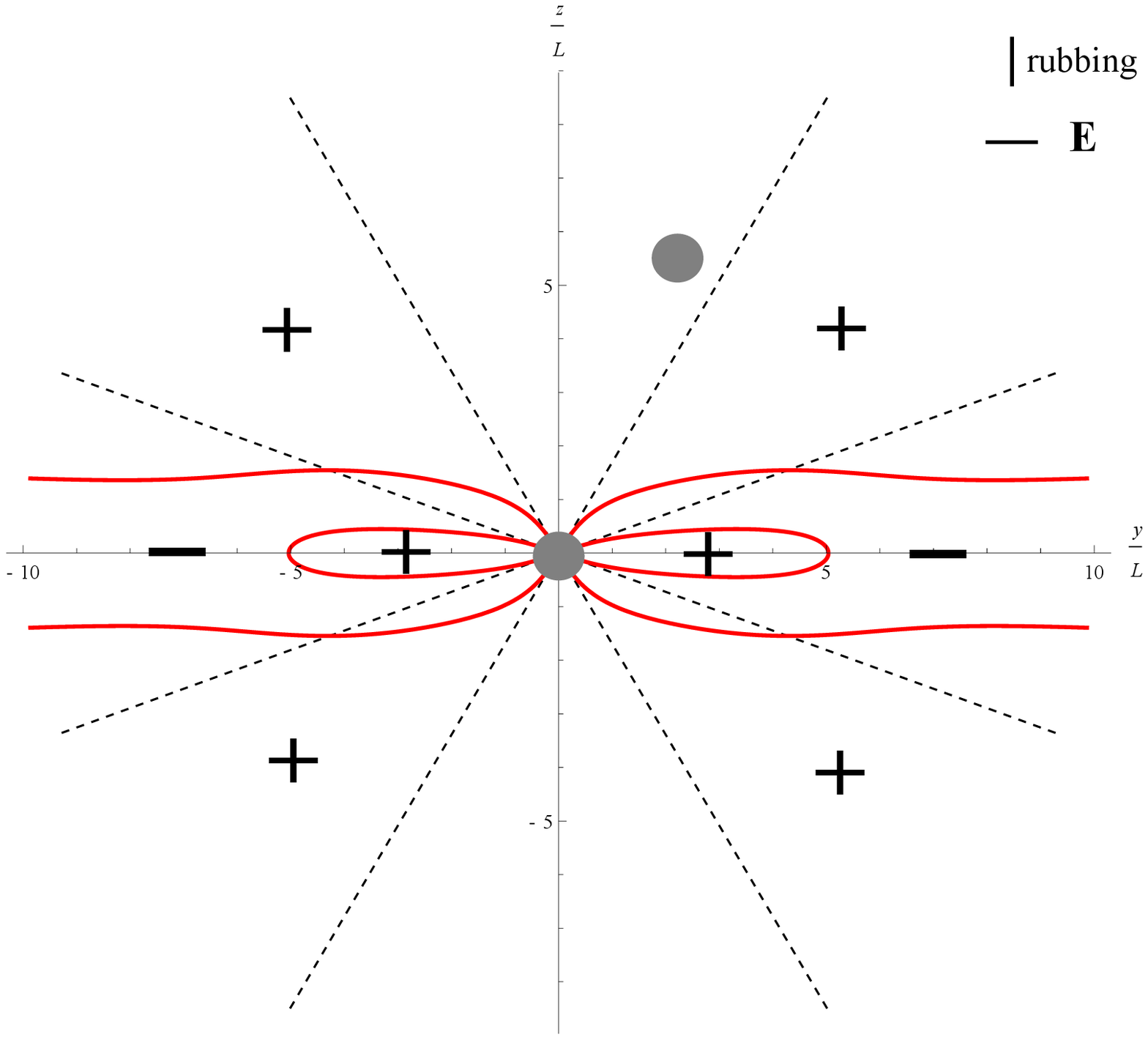}
\caption{(Color online) Map of the attraction and repulsion zones of quadrupole-quadrupole interaction between two particles in the middle of the planar cell with $\Delta\varepsilon<0$ placed in the electric field applied along the $y\text{-axis}, \textbf{E}||y$, $E=1.74E_{t}$. Dashed lines are zone's borders in the unlimited NLC. Collapse of the lateral tales into dumbbell-shaped region occurs at $E=1.74E_{t}$ (see Fig.\ref{qqbelow}). Sign ``-'' means attraction, ``+'' means repulsion, $cc^{\prime}>0$.}\label{qqabove}
\end{figure}

\begin{figure}
\includegraphics[clip=,width=\linewidth]{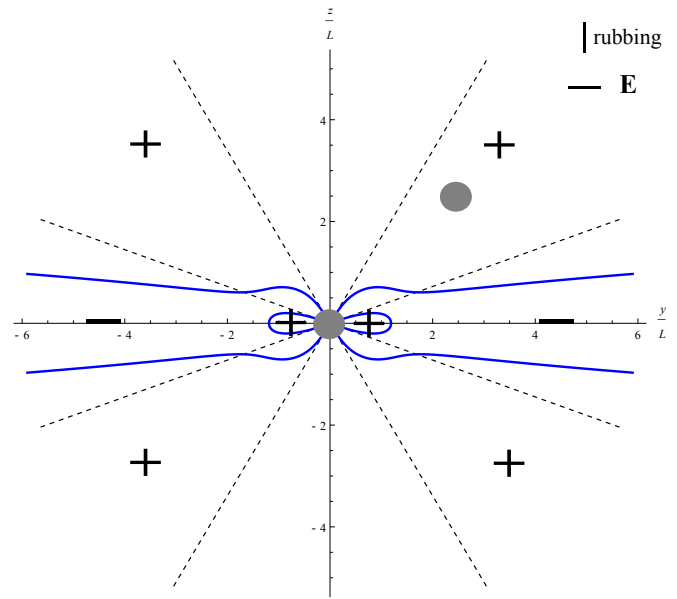}
\caption{(Color online) Map of the attraction and repulsion zones of quadrupole-quadrupole interaction between two particles in the middle of the planar cell with $\Delta\varepsilon<0$ placed in the electric field applied along the $y\text{-axis}, \textbf{E}||y$, $E=2E_{t}$. Dashed lines are zone's borders in the unlimited NLC. Compression of the dumbbell-shaped regions occurs with increasing of the field strength for $E>1.74E_{t}$ (see Fig.\ref{qqabove}). Sign ``-'' means attraction, ``+'' means repulsion, $cc^{\prime}>0$.}\label{qq2}
\end{figure}

\begin{figure*}
\begin{center}
\includegraphics[width=\textwidth]{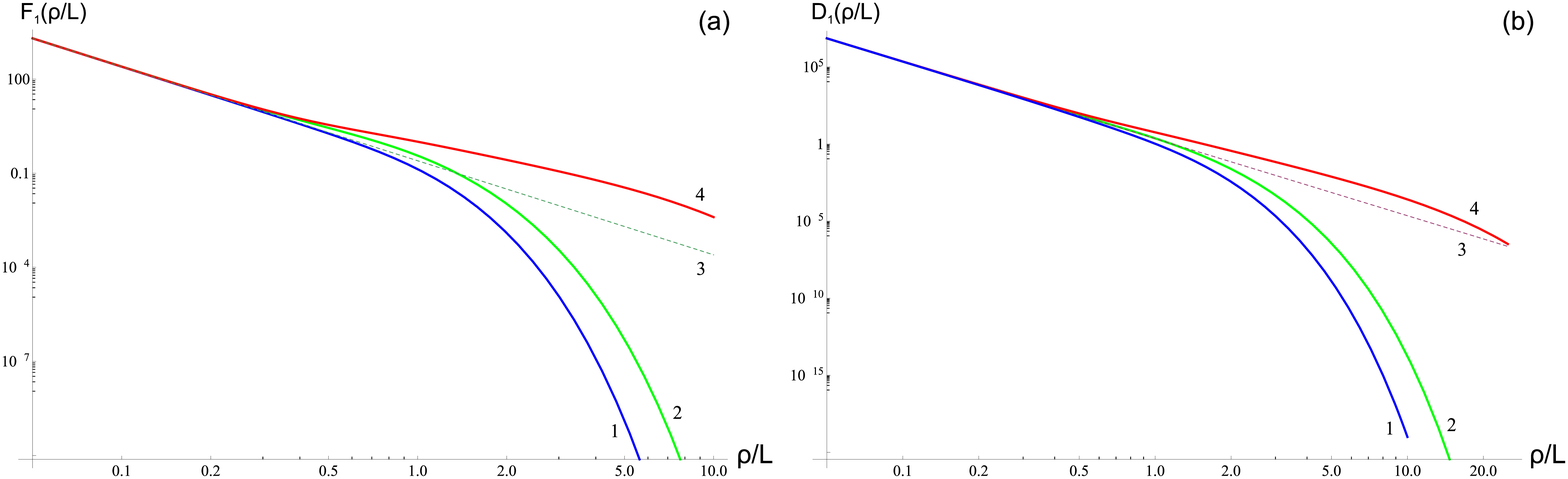}
\end{center}
\caption{(Color online) (a) Log-log plots of the $F_{1}$ as a function of the rescaled distance $\rho/L$. Dashed line 3 is the power-law asymptotics $\frac{1}{4}(\frac{L}{\rho})^{3}$. (b) Log-log plots of the $D_{1}$ as a function of the rescaled distance $\rho/L$. Dashed line 3 is the power-law asymptotics $\frac{9}{4}(\frac{L}{\rho})^{5}$. Planar cell, external electric field is applied along the $y\text{-axis}, \textbf{E}||y$ (see Fig.\ref{CS}b). Blue lines 1 correspond to $E=0.99E_{t}$ and $\Delta\varepsilon<0$. Green lines 2 correspond to the zero-field case $E=0$. Red lines 4 correspond to $E=0.99E_{t}$ and $\Delta\varepsilon>0$. This is a \textit{deconfinement effect} in the planar cell.}\label{Plan_y_log}
\end{figure*}

Let us consider the case with the field is parallel to the cell planes and perpendicular to the rubbing direction $\mathbf{n}_{0}$, i.e. $\textbf{E}||y$, $\mathbf{e}=(0,1,0)$. 

In order to get corresponding Green functions we can simply make a replacement $x \leftrightarrow y$ in \eqref{G_px}. One can easily find that such replacement means a replacement of $\lambda_{n}=\sqrt{\frac{n^2 \pi^2}{L^2}-k^2}$ with $\mu_{n}=\frac{n\pi}{L}$ and vice versa in \eqref{FF}. Formulas \eqref{U_dd_plan},\eqref{U_dQ_plan},\eqref{U_QQ_plan} are valid in this case as well as in the previous one. And their application to the interaction between two particles in the center of the cell $x=x^{\prime}=\frac{L}{2}$ is presented in Fig.\ref{Plan_y}. It is seen that the external field does not affect the interaction at small distances $\rho\ll L$. Though it changes maps of attraction and repulsion zones on far distances. Particularly, when the dielectric anisotropy is positive $\Delta\varepsilon>0$ (see Fig.\ref{Plan_y}, lines $a$), the electric field moves zone's borders closer to their analogues - cones - in the infinite nematic LC. This means decreasing of the cell planes confinement effect as electric field tends to rotate the director from $z$ direction to $y$ direction. This approaching $E\Rightarrow E_{t}$ produces the \textit{Deconfinment effect} in the planar cell similar to the deconfinement effect in the homeotropic cell described in Sec. III,a (see Fig. \ref{Plan_y_log}). And the Frederiks transition takes place at $E=E_{t}$ which breaks ground state $\textbf{n}_{0}||z$. 

More interesting changes of the interaction maps take place for negative dielectric anisotropy $\Delta\varepsilon<0$. We have no limitations on the field strength $E$ in this case. The electric field widens lateral tails of the dipole-dipole interaction with increasing of the strength value $E$ (see Fig. \ref{ddmany}). But there is the critical value $E_{col}=1.74E_{t}$ ($E_{t}=\frac{\pi}{L}\sqrt{\frac{4\pi K}{|\Delta\varepsilon|}}$) at which a collapse of the lateral tales into dumbbell-shaped region occurs (compare Fig. \ref{dfabove} and Fig. \ref{dfbelow}). Particles repel inside the dumbbell-shaped regions and attract outside of them (see Fig. \ref{dfabove}). So that the attraction in the perpendicular to $\textbf{n}_{0}$ direction occurs at $E>1.74E_{t}$.
Subsequent increasing of the field strength $E$ compresses dumbbell-shaped regions and makes perpendicular attraction more significant. The same situation takes place for quadrupole-quadrupole interaction (see Figs. \ref{qqbelow},\ref{qqabove},\ref{qq2}). It is interesting that the same collapse and origin of the dumbbell-shaped regions occurs at the same field strength $E_{col}=1.74E_{t}$ as for dipole-dipole interaction.

\subsection{Field parallel to the cell planes and parallel to the rubbing direction}

\begin{figure*}
\begin{center}
\includegraphics[width=\textwidth]{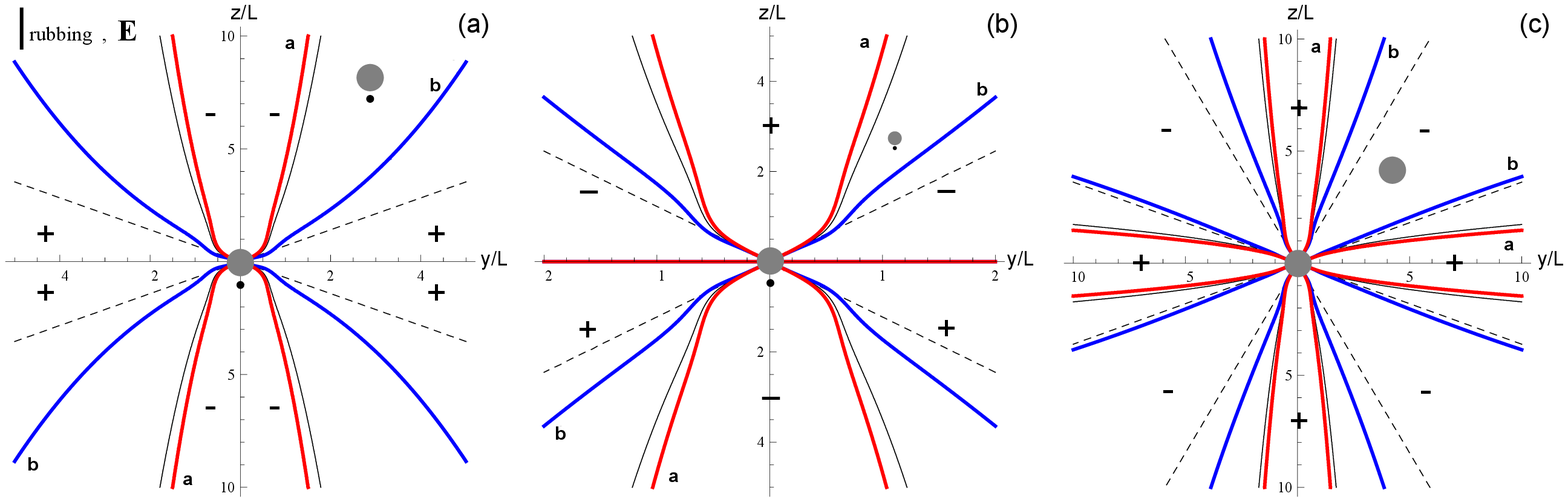}
\end{center}
\caption{(Color online) Map of the attraction and repulsion zones of interaction between two particles in the middle of the planar cell with rubbing along $z$ and thickness $L$  placed in the electric field parallel to $\mathbf{n}_{0}$, $E=0.99E_{t}$. Dashed lines are zone's borders in the unlimited NLC. Thick red lines $a$ correspond to the positive $\Delta\varepsilon>0$, thick blue lines $b$ correspond to the negative $\Delta\varepsilon<0$. Thin black lines are zone's borders in the zero-field case. Sign ``-'' means attraction, ``+'' means repulsion. (a) Dipole-dipole interaction, $p p^{\prime}>0$. (b) Dipole-quadrupole interaction. The larger particle is in the center, $pc^{\prime}>cp^{\prime}$. (c) Quadrupole-quadrupole interaction, $cc^{\prime}>0$.}\label{Plan_z}
\end{figure*}

Finally, let's consider the case when the electric field $\textbf{E}$ is parallel to the cell planes and parallel to the rubbing direction $\mathbf{n}_{0}$, i.e. $\textbf{E}||z$, $\mathbf{e}=(0,0,1)$. In this case $x$ and $y$ directions are equivalent and we can find the Green function from \eqref{G_px} as $G=G_{x}=G_{y}$ with $\mu_{n}=\lambda_{n}=\sqrt{\frac{n^2 \pi^2}{L^2}+k^2}$. Then all interactions are again given by formulas \eqref{U_dd_plan},\eqref{U_dQ_plan},\eqref{U_QQ_plan} with $\mu_{n}=\lambda_{n}=\sqrt{\frac{n^2 \pi^2}{L^2}+k^2}$. Maps of the interaction zones are depicted on Fig.\ref{Plan_z}. The electric field shrinks zones for positive dielectric anisotropy $\Delta\varepsilon>0$ and expands them to the unlimited nematic cones for negative $\Delta\varepsilon<0$. As in the Sec.III, A one can compare correspondent formulas with results of \cite{we} and find that an external field gives rise to a set of the effective cell thicknesses
\begin{equation}\label{L_eff}
L_{n}^{\text{eff}}=L \sqrt{\frac{n^2 E_{t}^2}{n^2 E_{t}^2+\text{sgn}(\Delta\varepsilon)E^2}}
\end{equation}
and thereby affects the interaction between particles. For zero field $E=0$ we have $L_{n}^{\text{eff}}=L$. It is clearly seen that when $\Delta\varepsilon>0$ subsequent increasing of the field strength $E$ will just make red zones on Fig.\ref{Plan_z} ad lib narrow as it just decreases the effective cell thickness $L_{n}\Rightarrow 0$ and increases screening strength.

 Of course there is the limitation $E\leq E_{t}$ and Frederiks transition takes place when $\Delta\varepsilon<0$. In this case as well $L_{1}^{\text{eff}}\rightarrow \infty$ as $E\rightarrow E_{t}$ so that the deconfinement effect takes place before the Frederiks transition point.

We see that maps of interactions on the Fig. \ref{Plan_z} are very resemble with Fig. \ref{dd_unlim_p}- Fig. \ref{QQ_unlim_p}. So we can try to extend our speculations in Sec.II for the planar nematic cell with the presence of the field. As we have already mentioned in the Section II, the external field applied in the NLC with positive $\Delta\varepsilon>0$  along the $\mathbf{n}_{0}$ induces the hedgehog transition to the Saturn-ring, i.e. dipole type defect transforms into quadrupole \cite{stark1,lopo,gu}. At the same time the $\Delta\varepsilon<0$ case leads to the deconfinement effect and the influence of the cell planes disappears near the critical point $E_{t}$ for dipole-dipole interaction. In some sense the confined NLC becomes similar to the unlimited nematic. But we know that in the unlimited nematic LC the hedgehog configuration is the most stable \cite{stark1}. So there should be the point where Saturn ring configuration loses it's stability, i.e. has higher energy than hedgehog. In the formulas \eqref{U_dd_plan},\eqref{U_dQ_plan},\eqref{U_QQ_plan} the main contribution has the term $n=1$, so we may \textit{put forward a hypothesis} that it's possible to replace $L$ with $L_{1}^{eff}$ in the (\ref{as}) and to estimate the point of stability of the Saturn ring configuration in the planar nematic cell with external electric field:

\begin{equation}
a_{S}\approx0.16 L_{1}^{eff}=\frac{0.16L}{\sqrt{1+\text{sgn}(\Delta\varepsilon)(\frac{E}{E_{t}})^{2}}}\label{asf}
\end{equation}
 For particles with radius $a>a_{S}$ the Saturn ring has the less energy than hedgehog and vice versa. When $\Delta\varepsilon>0$ the external field decreases effective cell width and strengthens Saturn ring stability. If the particle has radius $a<0.16L$ and has the hedgehog director configuration, the external field with value $E\geq E_{c}=E_{t}\sqrt{1+(\frac{0.16L}{a})^{2}}$,  ($E_{t}=\frac{\pi}{L}\sqrt{\frac{4\pi K}{|\Delta\varepsilon|}}$) should make Saturn ring configuration to be stable. We see that the threshold electric field $E_{c}$ decreases with increasing of the particle radius $a$. This tendency was experimentally observed in \cite{lopo} for strong anchoring strength at the particle's surface that is in line with our speculations.
 
But NLC with $\Delta\varepsilon<0$ has the opposite tendency. If the particle has radius $a>0.16L$ and has Saturn ring director configuration the external field with value $E\geq E_{c}=E_{t}\sqrt{1-(\frac{0.16L}{a})^{2}}$ makes hedgehog configuration to be stable.

 So it's possible to switch between Saturn ring and hedgehog configurations in the nematic cell with help of the electric (magnetic) field. Of course our speculations here are just \textit{a hypothesis} and there should be experimental or computer simulation test to check quantitatively the relation (\ref{asf}).

\section{Conclusions} 

In this paper we propose the Green function method which enables to describe quantitatively elastic colloidal interactions between axially symmetric particles confined in nematic cell under the action of the external electric (magnetic) fields. The current method is the generalization of the method proposed in \cite{we,we2} for description of the colloids in confined NLC. General formulas for dipole-dipole, dipole-quadrupole and quadrupole-quadrupole interactions in the homeotropic and planar nematic cells with parallel and perpendicular field orientations are obtained. We consider cases of homeotropic as well as planar nematic cells and have found some new results: 

1)We propose \textit{a criterion} \eqref{as} between radius of the particle and cell thickness when the Saturn ring configuration becomes stable ( for particle radius more than $a_{S}$) in the nematic cell in comparison with point hedgehog configuration and generalize this criterion on the case of the applied field \eqref{asf} in the planar cell. Of course this criterion is not an exact result but rather an estimate which comes from analogy between exact results \eqref{dd_unl}-\eqref{QQ_unl} for unlimited NLC with external field and results for confined NLC with- and without external field. It should be tested with computer simulations or experiments.

2) \textit{Deconfinement effect} for dipole particles in the homeotropic nematic cell with negative dielectric  anisotropy $\Delta\varepsilon<0$ and perpendicular to the cell electric field when electric field is approaching it's Frederiks threshold value $E\Rightarrow E_{c}$. This means cancellation of the confinement effect for dipole particles (in \cite{conf} the confinement effect was found for quadrupole particles)  near the Frederiks transition while it remains for quadrupole particles. The same effect takes place in the planar cell with $\Delta\varepsilon>0$, $\textbf{E}\bot \textbf{n}_{0}$ and in the planar cell with $\Delta\varepsilon<0$, $\textbf{E}||\textbf{n}_{0}$.

3) New effect of \textit{attraction and stabilization} of the particles in the homeotropic nematic cell along the electric field parallel to the cell planes when nematic dielectric anisotropy is negative $\Delta\varepsilon<0$. The minimun distance between two particles depends on the strength of the field and can be ordinary for $\Delta\varepsilon<0$ (see Fig. \ref{rho}).

 4) Attraction and repulsion zones for all elastic interactions are changed dramatically under the action of the external field. Especially when the electric field $\textbf{E}$ is parallel to the planar cell planes and perpendicular to the rubbing direction $\mathbf{n}_{0}$ and NLC has negative dielectric anisotropy $\Delta\varepsilon<0$. The electric field widens lateral tails of the dipole-dipole interaction with increasing of the strength value $E$ (see Fig. \ref{ddmany}). But there is the critical value $E_{col}=1.74E_{t}$ ($E_{t}=\frac{\pi}{L}\sqrt{\frac{4\pi K}{|\Delta\varepsilon|}}$) at which a collapse of the lateral tales into dumbbell-shaped region occurs (compare Fig. \ref{dfbelow} and Fig. \ref{dfabove} as well as Fig. \ref{qqbelow} and Fig. \ref{qqabove}). Particles repel inside the dumbbell-shaped regions and attract outside of them (see Fig. \ref{dfabove}). So that the attraction in the perpendicular to $\textbf{n}_{0}$ direction occurs at $E>E_{col}=1.74E_{t}$. Subsequent increasing of the field strength $E$ compresses dumbbell-shaped regions and makes perpendicular attraction more significant. The same situation takes place for quadrupole-quadrupole interaction (see Figs. \ref{qqbelow},\ref{qqabove},\ref{qq2}). It is interesting that the same collapse and origin of the dumbbell-shaped regions occurs at the same field strength $E_{col}=1.74E_{t}$ as for the dipole-dipole interaction. 
 
We hope that listed above results may be tested either with help of computer simulations or experimentally and would be helpful in further investigations of fascinating world of colloids in liquid crystals.

\end{document}